%
%
%
\def\unredoffs{} \def\redoffs{\voffset=-.31truein\hoffset=-.59truein}
\def\speclscape{\special{ps: landscape}}
%
%
%
%
\newbox\leftpage \newdimen\fullhsize \newdimen\hstitle \newdimen\hsbody
\tolerance=1000\hfuzz=2pt
\catcode`\@=11 
\def\bigans{b }
\def\answ{b }
\ifx\answ\bigans\message{(This will come out unreduced.}
\magnification=1200\unredoffs\baselineskip=16pt plus 2pt minus 1pt
\hsbody=\hsize \hstitle=\hsize 
\else\message{(This will be reduced.} \let\l@r=L
\magnification=1000\baselineskip=16pt plus 2pt minus 1pt \vsize=7truein
\redoffs \hstitle=8truein\hsbody=4.75truein\fullhsize=10truein\hsize=\hsbody
\output={\ifnum\pageno=0 
  \shipout\vbox{\speclscape{\hsize\fullhsize\makeheadline}
    \hbox to \fullhsize{\hfill\pagebody\hfill}}\advancepageno
  \else
  \almostshipout{\leftline{\vbox{\pagebody\makefootline}}}\advancepageno
  \fi}
\def\almostshipout#1{\if L\l@r \count1=1 \message{[\the\count0.\the\count1]}
      \global\setbox\leftpage=#1 \global\let\l@r=R
 \else \count1=2
  \shipout\vbox{\speclscape{\hsize\fullhsize\makeheadline}
      \hbox to\fullhsize{\box\leftpage\hfil#1}}  \global\let\l@r=L\fi}
\fi
%
\newcount\yearltd\yearltd=\year\advance\yearltd by -1900

\def\Title#1#2{\nopagenumbers\abstractfont\hsize=\hstitle\rightline{#1}%
\vskip 1in\centerline{\titlefont #2}\abstractfont\vskip .5in\pageno=0}
\def\Date#1{\vfill\leftline{#1}\tenpoint\supereject\global\hsize=\hsbody%
\footline={\hss\tenrm\folio\hss}}
%

\def\draftmode{\message{ DRAFTMODE }\def\draftdate{{\rm preliminary draft:
\number\month/\number\day/\number\yearltd\ \ \hourmin}}%
\headline={\hfil\draftdate}\writelabels\baselineskip=20pt plus 2pt minus 2pt
 {\count255=\time\divide\count255 by 60 \xdef\hourmin{\number\count255}
  \multiply\count255 by-60\advance\count255 by\time
  \xdef\hourmin{\hourmin:\ifnum\count255<10 0\fi\the\count255}}}
\def\nolabels{\def\wrlabeL##1{}\def\eqlabeL##1{}\def\reflabeL##1{}}
\def\writelabels{\def\wrlabeL##1{\leavevmode\vadjust{\rlap{\smash%
{\line{{\escapechar=` \hfill\rlap{\sevenrm\hskip.03in\string##1}}}}}}}%
\def\eqlabeL##1{{\escapechar-1\rlap{\sevenrm\hskip.05in\string##1}}}%
\def\reflabeL##1{\noexpand\llap{\noexpand\sevenrm\string\string\string##1}}}
\nolabels
%
\global\newcount\secno \global\secno=0
\global\newcount\meqno \global\meqno=1
\def\newsec#1{\global\advance\secno by1\message{(\the\secno. #1)}
\global\subsecno=0\eqnres@t\noindent{\bf\the\secno. #1}
\writetoca{{\secsym} {#1}}\par\nobreak\medskip\nobreak}
\def\eqnres@t{\xdef\secsym{\the\secno.}\global\meqno=1\bigbreak\bigskip}
\def\sequentialequations{\def\eqnres@t{\bigbreak}}\xdef\secsym{}
\global\newcount\subsecno \global\subsecno=0
\def\subsec#1{\global\advance\subsecno by1\message{(\secsym\the\subsecno.
#1)}
\ifnum\lastpenalty>9000\else\bigbreak\fi
\noindent{\it\secsym\the\subsecno. #1}\writetoca{\string\quad
{\secsym\the\subsecno.} {#1}}\par\nobreak\medskip\nobreak}
\def\appendix#1#2{\global\meqno=1\global\subsecno=0\xdef\secsym{\hbox{#1.}}
\bigbreak\bigskip\noindent{\bf Appendix #1. #2}\message{(#1. #2)}
\writetoca{Appendix {#1.} {#2}}\par\nobreak\medskip\nobreak}
%
%
\def\eqnn#1{\xdef #1{(\secsym\the\meqno)}\writedef{#1\leftbracket#1}%
\global\advance\meqno by1\wrlabeL#1}
\def\eqna#1{\xdef #1##1{\hbox{$(\secsym\the\meqno##1)$}}
\writedef{#1\numbersign1\leftbracket#1{\numbersign1}}%
\global\advance\meqno by1\wrlabeL{#1$\{\}$}}
\def\eqn#1#2{\xdef #1{(\secsym\the\meqno)}\writedef{#1\leftbracket#1}%
\global\advance\meqno by1$$#2\eqno#1\eqlabeL#1$$}
%
\newskip\footskip\footskip14pt plus 1pt minus 1pt 
\def\footnotefont{\ninepoint}\def\f@t#1{\footnotefont #1\@foot}
\def\f@@t{\baselineskip\footskip\bgroup\footnotefont\aftergroup\@foot\let\next}
\setbox\strutbox=\hbox{\vrule height9.5pt depth4.5pt width0pt}
\global\newcount\ftno \global\ftno=0
\def\foot{\global\advance\ftno by1\footnote{$^{\the\ftno}$}}
%
\newwrite\ftfile
\def\footend{\def\foot{\global\advance\ftno by1\chardef\wfile=\ftfile
$^{\the\ftno}$\ifnum\ftno=1\immediate\openout\ftfile=foots.tmp\fi%
\immediate\write\ftfile{\noexpand\smallskip%
\noexpand\item{f\the\ftno:\ }\pctsign}\findarg}%
\def\footatend{\vfill\eject\immediate\closeout\ftfile{\parindent=20pt
\centerline{\bf Footnotes}\nobreak\bigskip\input foots.tmp }}}
\def\footatend{}
%
%
\global\newcount\refno \global\refno=1
\newwrite\rfile
\def\ref{[\the\refno]\nref}
\def\nref#1{\xdef#1{[\the\refno]}\writedef{#1\leftbracket#1}%
\ifnum\refno=1\immediate\openout\rfile=refs.tmp\fi
\global\advance\refno by1\chardef\wfile=\rfile\immediate
\write\rfile{\noexpand\item{#1\ }\reflabeL{#1\hskip.31in}\pctsign}\findarg}
\def\findarg#1#{\begingroup\obeylines\newlinechar=`\^^M\pass@rg}
{\obeylines\gdef\pass@rg#1{\writ@line\relax #1^^M\hbox{}^^M}%
\gdef\writ@line#1^^M{\expandafter\toks0\expandafter{\striprel@x #1}%
\edef\next{\the\toks0}\ifx\next\em@rk\let\next=\endgroup\else\ifx\next\empty%
\else\immediate\write\wfile{\the\toks0}\fi\let\next=\writ@line\fi\next\relax}}
\def\striprel@x#1{} \def\em@rk{\hbox{}}
\def\lref{\begingroup\obeylines\lr@f}
\def\lr@f#1#2{\gdef#1{\ref#1{#2}}\endgroup\unskip}
\def\semi{;\hfil\break}
\def\addref#1{\immediate\write\rfile{\noexpand\item{}#1}} 
\def\startrefs#1{\immediate\openout\rfile=refs.tmp\refno=#1}
\def\xref{\expandafter\xr@f}\def\xr@f[#1]{#1}
\def\refs#1{\count255=1[\r@fs #1{\hbox{}}]}
\def\r@fs#1{\ifx\und@fined#1\message{reflabel \string#1 is undefined.}%
\nref#1{need to supply reference \string#1.}\fi%
\vphantom{\hphantom{#1}}\edef\next{#1}\ifx\next\em@rk\def\next{}%
\else\ifx\next#1\ifodd\count255\relax\xref#1\count255=0\fi%
\else#1\count255=1\fi\let\next=\r@fs\fi\next}
%

%
\newwrite\ffile\global\newcount\figno \global\figno=1
\def\fig{fig.~\the\figno\nfig}
\def\nfig#1{\xdef#1{fig.~\the\figno}%
\writedef{#1\leftbracket fig.\noexpand~\the\figno}%
\ifnum\figno=1\immediate\openout\ffile=figs.tmp\fi\chardef\wfile=\ffile%
\immediate\write\ffile{\noexpand\medskip\noexpand\item{Fig.\ \the\figno. }
\reflabeL{#1\hskip.55in}\pctsign}\global\advance\figno by1\findarg}
\def\xfig{\expandafter\xf@g}\def\xf@g fig.\penalty\@M\ {}
\def\figs#1{figs.~\f@gs #1{\hbox{}}}
\def\f@gs#1{\edef\next{#1}\ifx\next\em@rk\def\next{}\else
\ifx\next#1\xfig #1\else#1\fi\let\next=\f@gs\fi\next}
\newwrite\lfile
{\escapechar-1\xdef\pctsign{\string\%}\xdef\leftbracket{\string\{}
\xdef\rightbracket{\string\}}\xdef\numbersign{\string\#}}

\def\writestop{\def\writestoppt{\immediate\write\lfile{\string\pageno%
\the\pageno\string\startrefs\leftbracket\the\refno\rightbracket%
\string\def\string\secsym\leftbracket\secsym\rightbracket%
\string\secno\the\secno\string\meqno\the\meqno}\immediate\closeout\lfile}}
\def\writestoppt{}\def\writedef#1{}
\def\seclab#1{\xdef #1{\the\secno}\writedef{#1\leftbracket#1}\wrlabeL{#1=#1}}
\def\subseclab#1{\xdef #1{\secsym\the\subsecno}%
\writedef{#1\leftbracket#1}\wrlabeL{#1=#1}}
\newwrite\tfile \def\writetoca#1{}
\def\leaderfill{\leaders\hbox to 1em{\hss.\hss}\hfill}
\def\writetoc{\immediate\openout\tfile=toc.tmp
   \def\writetoca##1{{\edef\next{\write\tfile{\noindent ##1
   \string\leaderfill {\noexpand\number\pageno} \par}}\next}}}

%
\catcode`\@=12 
%
\edef\tfontsize{\ifx\answ\bigans scaled\magstep3\else scaled\magstep4\fi}
\font\titlerm=cmr10 \tfontsize \font\titlerms=cmr7 \tfontsize
\font\titlermss=cmr5 \tfontsize \font\titlei=cmmi10 \tfontsize
\font\titleis=cmmi7 \tfontsize \font\titleiss=cmmi5 \tfontsize
\font\titlesy=cmsy10 \tfontsize \font\titlesys=cmsy7 \tfontsize
\font\titlesyss=cmsy5 \tfontsize \font\titleit=cmti10 \tfontsize
\skewchar\titlei='177 \skewchar\titleis='177 \skewchar\titleiss='177
\skewchar\titlesy='60 \skewchar\titlesys='60 \skewchar\titlesyss='60
\def\titlefont{\def\rm{\fam0\titlerm}
\textfont0=\titlerm \scriptfont0=\titlerms \scriptscriptfont0=\titlermss
\textfont1=\titlei \scriptfont1=\titleis \scriptscriptfont1=\titleiss
\textfont2=\titlesy \scriptfont2=\titlesys \scriptscriptfont2=\titlesyss
\textfont\itfam=\titleit \def\it{\fam\itfam\titleit}\rm}
 \ifx\answ\bigans\else scaled\magstep1\fi
\ifx\answ\bigans\def\abstractfont{\tenpoint}\else
\font\abssl=cmsl10 scaled \magstep1
\font\absrm=cmr10 scaled\magstep1 \font\absrms=cmr7 scaled\magstep1
\font\absrmss=cmr5 scaled\magstep1 \font\absi=cmmi10 scaled\magstep1
\font\absis=cmmi7 scaled\magstep1 \font\absiss=cmmi5 scaled\magstep1
\font\abssy=cmsy10 scaled\magstep1 \font\abssys=cmsy7 scaled\magstep1
\font\abssyss=cmsy5 scaled\magstep1 \font\absbf=cmbx10 scaled\magstep1
\skewchar\absi='177 \skewchar\absis='177 \skewchar\absiss='177
\skewchar\abssy='60 \skewchar\abssys='60 \skewchar\abssyss='60
\def\abstractfont{\def\rm{\fam0\absrm}
\textfont0=\absrm \scriptfont0=\absrms \scriptscriptfont0=\absrmss
\textfont1=\absi \scriptfont1=\absis \scriptscriptfont1=\absiss
\textfont2=\abssy \scriptfont2=\abssys \scriptscriptfont2=\abssyss
\textfont\itfam=\bigit \def\it{\fam\itfam\bigit}\def\footnotefont{\tenpoint}%
\textfont\slfam=\abssl \def\sl{\fam\slfam\abssl}%
\textfont\bffam=\absbf \def\bf{\fam\bffam\absbf}\rm}\fi
\def\tenpoint{\def\rm{\fam0\tenrm}
\textfont0=\tenrm \scriptfont0=\sevenrm \scriptscriptfont0=\fiverm
\textfont1=\teni  \scriptfont1=\seveni  \scriptscriptfont1=\fivei
\textfont2=\tensy \scriptfont2=\sevensy \scriptscriptfont2=\fivesy
\textfont\itfam=\tenit
\def\it{\fam\itfam\tenit}\def\footnotefont{\ninepoint}%
\textfont\bffam=\tenbf \def\bf{\fam\bffam\tenbf}\def\sl{\fam\slfam\tensl}\rm}
\font\ninerm=cmr9 \font\sixrm=cmr6 \font\ninei=cmmi9 \font\sixi=cmmi6
\font\ninesy=cmsy9 \font\sixsy=cmsy6 \font\ninebf=cmbx9
\font\nineit=cmti9 \font\ninesl=cmsl9 \skewchar\ninei='177
\skewchar\sixi='177 \skewchar\ninesy='60 \skewchar\sixsy='60
\def\ninepoint{\def\rm{\fam0\ninerm}
\textfont0=\ninerm \scriptfont0=\sixrm \scriptscriptfont0=\fiverm
\textfont1=\ninei \scriptfont1=\sixi \scriptscriptfont1=\fivei
\textfont2=\ninesy \scriptfont2=\sixsy \scriptscriptfont2=\fivesy
\textfont\itfam=\ninei \def\it{\fam\itfam\nineit}\def\sl{\fam\slfam\ninesl}%
\textfont\bffam=\ninebf \def\bf{\fam\bffam\ninebf}\rm}
%
%

\hyphenation{anom-aly anom-alies coun-ter-term coun-ter-terms}
\def\inv{^{\raise.15ex\hbox{${\scriptscriptstyle -}$}\kern-.05em 1}}

\def\Dsl{\,\raise.15ex\hbox{/}\mkern-13.5mu D} 
\def\dsl{\raise.15ex\hbox{/}\kern-.57em\partial}

 \def\Tr{{\rm Tr}}
\font\bigit=cmti10 scaled \magstep1
\def\lspace{\ifx\answ\bigans{}\else\qquad\fi}
\def\lbspace{\ifx\answ\bigans{}\else\hskip-.2in\fi} 
\def\boxeqn#1{\vcenter{\vbox{\hrule\hbox{\vrule\kern3pt\vbox{\kern3pt
           \hbox{${\displaystyle #1}$}\kern3pt}\kern3pt\vrule}\hrule}}}
\def\mbox#1#2{\vcenter{\hrule \hbox{\vrule height#2in
               \kern#1in \vrule} \hrule}}  
%
  \def\CF{{\cal F}} 
\def\CL{{\cal L}}  \def\CI{{\cal I}} 
   \def\CT{{\cal T}}
\def\e#1{{\rm e}^{^{\textstyle#1}}}

\def\darr#1{\raise1.5ex\hbox{$\leftrightarrow$}\mkern-16.5mu #1}

\def\roughly#1{\raise.3ex\hbox{$#1$\kern-.75em\lower1ex\hbox{$\sim$}}}



\def\IB{\relax\hbox{$\inbar\kern-.3em{\rm B}$}}
\def\IC{\relax\hbox{$\inbar\kern-.3em{\rm C}$}}
\def\ID{\relax\hbox{$\inbar\kern-.3em{\rm D}$}}
\def\IE{\relax\hbox{$\inbar\kern-.3em{\rm E}$}}
\def\IF{\relax\hbox{$\inbar\kern-.3em{\rm F}$}}
\def\IG{\relax\hbox{$\inbar\kern-.3em{\rm G}$}}
\def\IGa{\relax\hbox{${\rm I}\kern-.18em\Gamma$}}
\def\IH{\relax{\rm I\kern-.18em H}}
\def\IK{\relax{\rm I\kern-.18em K}}
\def\II{\relax{\rm I\kern-.18em I}}
\def\IL{\relax{\rm I\kern-.18em L}}
\def\IP{\relax{\rm I\kern-.18em P}}
\def\IR{\relax{\rm I\kern-.18em R}}
\def\IZ{\relax\ifmmode\mathchoice {\hbox{\cmss Z\kern-.4em Z}}{\hbox{\cmss
Z\kern-.4em Z}} {\lower.9pt\hbox{\cmsss Z\kern-.4em Z}}
{\lower1.2pt\hbox{\cmsss Z\kern-.4em Z}}\else{\cmss Z\kern-.4em Z}\fi}

\def\IB{\relax{\rm I\kern-.18em B}}
\def\IC{{\relax\hbox{$\inbar\kern-.3em{\rm C}$}}}
\def\ID{\relax{\rm I\kern-.18em D}}
\def\IE{\relax{\rm I\kern-.18em E}}
\def\IF{\relax{\rm I\kern-.18em F}}


\def\CF {{\cal F}}

\def\CI {{\cal I}}

\def\CL {{\cal L}}

\def\CN {{\cal N}}

\def\CT {{\cal T}}

\def\CW {{\cal W}}

\def\p{\partial}
\def\pa{\partial}
\def\pb{{\bar{\partial}}}




\def\s{\lies}

\def\Tr{{\rm Tr}}


\def\demi{{1\over 2}}

\def\c{\cdot}


\def\f{\phi}    
\def\P{\Psi}    
\def\F{\Phi}

\def\a{\alpha}
\def\b{\beta}
\def\g{\gamma}  
\def\d{\delta}  
\def\m{\mu}
\def\n{\nu}
\def\r{\rho}
\def\l{\lambda} \def\L{\Lambda}

\def\e{\epsilon}

\def\|{\Big|}
\def\({\Big(}   \def\){\Big)}
\def\[{\Big[}   \def\]{\Big]}



\def\paper#1#2#3#4{#1, {\sl #2}, #3 {\tt #4}}

\def\hh{hep-th/}


\def\PLB#1#2#3{Phys. Lett.~{\bf B#1} (#2) #3}
\def\NPB#1#2#3{Nucl. Phys.~{\bf B#1} (#2) #3}
\def\PRL#1#2#3{Phys. Rev. Lett.~{\bf #1} (#2) #3}
\def\CMP#1#2#3{Comm. Math. Phys.~{\bf #1} (#2) #3}
\def\PRD#1#2#3{Phys. Rev.~{\bf D#1} (#2) #3}
\def\MPL#1#2#3{Mod. Phys. Lett.~{\bf #1} (#2) #3}
\def\IJMP#1#2#3{Int. Jour. Mod. Phys.~{\bf #1} (#2) #3}


\def\unlockat{\catcode`\@=11}
\def\lockat{\catcode`\@=12}

\unlockat


\def\newsec#1{\global\advance\secno by1\message{(\the\secno. #1)}
\global\subsecno=0\global\subsubsecno=0\eqnres@t\noindent {\bf\the\secno. #1}
\writetoca{{\secsym} {#1}}\par\nobreak\medskip\nobreak}
\global\newcount\subsecno \global\subsecno=0
\def\subsec#1{\global\advance\subsecno by1\message{(\secsym\the\subsecno.
#1)}
\ifnum\lastpenalty>9000\else\bigbreak\fi\global\subsubsecno=0
\noindent{\it\secsym\the\subsecno. #1}
\writetoca{\string\quad {\secsym\the\subsecno.} {#1}}
\par\nobreak\medskip\nobreak}
\global\newcount\subsubsecno \global\subsubsecno=0
\def\subsubsec#1{\global\advance\subsubsecno by1
\message{(\secsym\the\subsecno.\the\subsubsecno. #1)}
\ifnum\lastpenalty>9000\else\bigbreak\fi
\noindent\quad{\secsym\the\subsecno.\the\subsubsecno.}{#1}
\writetoca{\string\qquad{\secsym\the\subsecno.\the\subsubsecno.}{#1}}
\par\nobreak\medskip\nobreak}

\def\subsubseclab#1{\DefWarn#1\xdef #1{\noexpand\hyperref{}{subsubsection}%
{\secsym\the\subsecno.\the\subsubsecno}%
{\secsym\the\subsecno.\the\subsubsecno}}%
\writedef{#1\leftbracket#1}\wrlabeL{#1=#1}}
\lockat

\def\dbend{\lower3.5pt\hbox{\manual\char127}}


\def\boxit#1{\vbox{\hrule\hbox{\vrule\kern8pt
\vbox{\hbox{\kern8pt}\hbox{\vbox{#1}}\hbox{\kern8pt}}
\kern8pt\vrule}\hrule}}

\def\mathboxit#1{\vbox{\hrule\hbox{\vrule\kern8pt\vbox{\kern8pt
\hbox{$\displaystyle #1$}\kern8pt}\kern8pt\vrule}\hrule}}


\def\inbar{\,\vrule height1.5ex width.4pt depth0pt}

\font\cmss=cmss10 \font\cmsss=cmss10 at 7pt


\lref\simons{ J. Cheeger and J. Simons, {\it Differential Characters and
Geometric Invariants},  Stony Brook Preprint, (1973), unpublished.}

\lref\cargese{ L.~Baulieu, {\it Algebraic quantization of gauge theories},   
Perspectives in fields and particles, Plenum Press, eds. Basdevant-Levy,
Cargese Lectures 1983}

\lref\antifields{ L. Baulieu, M. Bellon, S. Ouvry, C.Wallet, Phys.Letters
B252 (1990) 387; M.  Bocchichio, Phys. Lett. B187 (1987) 322;  Phys. Lett. B
192 (1987) 31; R.  Thorn    Nucl. Phys.   B257 (1987) 61. }

\lref\thompson{ George Thompson,  Annals Phys. 205 (1991) 130; J.M.F.
Labastida, M. Pernici, Phys. Lett. 212B  (1988) 56; D. Birmingham, M.Blau,
M. Rakowski and G.Thompson, Phys. Rept. 209 (1991) 129.}

\lref\tonin{ Tonin}

\lref\wittensix{ E.  Witten, {\it New  Gauge  Theories In Six Dimensions},
\hh{9710065}. }

\lref\orlando{ O. Alvarez, L. A. Ferreira and J. Sanchez Guillen, {\it  A New
Approach to Integrable Theories in any Dimension}, hep-th/9710147.}

\lref\wittentopo{ E.  Witten,  {\it  Topological Quantum Field Theory},
\hh9403195, Commun.  Math. Phys.  {117} (1988)353.  }

\lref\wittentwist{ E.  Witten, {\it Supersymmetric Yang--Mills theory on a
four-manifold}, J.  Math.  Phys.  {35} (1994) 5101.}

\lref\west{ L.~Baulieu, P.~West, {\it Six Dimensional TQFTs and  Self-dual
Two-Forms,} Phys.Lett. B {\bf 436 } (1998) 97, /hep-th/9805200}

\lref\bv{ I.A. Batalin and V.A. Vilkowisky,    Phys. Rev.   D28  (1983)
2567\semi M. Henneaux,  Phys. Rep.  126   (1985) 1\semi M. Henneaux and C.
Teitelboim, {\it Quantization of Gauge Systems}
  Princeton University Press,  Princeton (1992).}

\lref\kyoto{ L. Baulieu, E. Bergschoeff and E. Sezgin, Nucl. Phys.
B307(1988)348\semi L. Baulieu,   {\it Field Antifield Duality, p-Form Gauge
Fields
   and Topological Quantum Field Theories}, hep-th/9512026,
   Nucl. Phys. B478 (1996) 431.  }

\lref\sourlas{ G. Parisi and N. Sourlas, {\it Random Magnetic Fields,
Supersymmetry and Negative Dimensions}, Phys. Rev. Lett.  43 (1979) 744;
Nucl.  Phys.  B206 (1982) 321.  }

\lref\SalamSezgin{ A.  Salam  and  E.  Sezgin, {\it Supergravities in
diverse dimensions}, vol.  1, p. 119\semi P.  Howe, G.  Sierra and P. 
Townsend, Nucl Phys B221 (1983) 331.}

\lref\nekrasov{ A. Losev, G. Moore, N. Nekrasov, S. Shatashvili, {\it
Four-Dimensional Avatars of Two-Dimensional RCFT},  hep-th/9509151, Nucl. 
Phys.  Proc.  Suppl.   46 (1996) 130\semi L.  Baulieu, A.  Losev,
N.~Nekrasov  {\it Chern-Simons and Twisted Supersymmetry in Higher
Dimensions},  hep-th/9707174, to appear in Nucl.  Phys.  B.  }

\lref\WitDonagi{R.~ Donagi, E.~ Witten, ``Supersymmetric Yang--Mills Theory
and Integrable Systems'', hep-th/9510101, Nucl. Phys.{\bf B}460 (1996)
299-334}
\lref\Witfeb{E.~ Witten, ``Supersymmetric Yang--Mills Theory On A
Four-Manifold,''  hep-th/9403195; J. Math. Phys. {\bf 35} (1994) 5101.}
\lref\Witgrav{E.~ Witten, ``Topological Gravity'', Phys.Lett.206B:601, 1988}
\lref\witaffl{I. ~ Affleck, J.A.~ Harvey and E.~ Witten,
        ``Instantons and (Super)Symmetry Breaking
        in $2+1$ Dimensions'', Nucl. Phys. {\bf B}206 (1982) 413}
\lref\wittabl{E.~ Witten,  ``On $S$-Duality in Abelian Gauge Theory,''
hep-th/9505186; Selecta Mathematica {\bf 1} (1995) 383}
\lref\wittgr{E.~ Witten, ``The Verlinde Algebra And The Cohomology Of The
Grassmannian'',  hep-th/9312104}
\lref\wittenwzw{E. Witten, ``Non abelian bosonization in two dimensions,''
Commun. Math. Phys. {\bf 92} (1984)455 }
\lref\witgrsm{E. Witten, ``Quantum field theory, grassmannians and algebraic
curves,'' Commun.Math.Phys.113:529,1988}
\lref\wittjones{E. Witten, ``Quantum field theory and the Jones
polynomial,'' Commun.  Math. Phys., 121 (1989) 351. }
\lref\witttft{E.~ Witten, ``Topological Quantum Field Theory", Commun. Math.
Phys. {\bf 117} (1988) 353.}
\lref\wittmon{E.~ Witten, ``Monopoles and Four-Manifolds'', hep-th/9411102}
\lref\Witdgt{ E.~ Witten, ``On Quantum gauge theories in two dimensions,''
Commun. Math. Phys. {\bf  141}  (1991) 153}
\lref\witrevis{E.~ Witten,
 ``Two-dimensional gauge theories revisited'', hep-th/9204083; J. Geom.
Phys. 9 (1992) 303-368}
\lref\Witgenus{E.~ Witten, ``Elliptic Genera and Quantum Field Theory'',
Comm. Math. Phys. 109(1987) 525. }
\lref\OldZT{E. Witten, ``New Issues in Manifolds of SU(3) Holonomy,'' {\it
Nucl. Phys.} {\bf B268} (1986) 79 \semi J. Distler and B. Greene, ``Aspects
of (2,0) String Compactifications,'' {\it Nucl. Phys.} {\bf B304} (1988) 1
\semi B. Greene, ``Superconformal Compactifications in Weighted Projective
Space,'' {\it Comm. Math. Phys.} {\bf 130} (1990) 335.}
\lref\bagger{E.~ Witten and J. Bagger, Phys. Lett. {\bf 115B}(1982) 202}
\lref\witcurrent{E.~ Witten,``Global Aspects of Current Algebra'',
Nucl.Phys.B223 (1983) 422\semi ``Current Algebra, Baryons and Quark
Confinement'', Nucl.Phys. B223 (1993) 433}
\lref\Wittreiman{S.B. Treiman, E. Witten, R. Jackiw, B. Zumino, ``Current
Algebra and Anomalies'', Singapore, Singapore: World Scientific ( 1985) }
\lref\Witgravanom{L. Alvarez-Gaume, E.~ Witten, ``Gravitational Anomalies'',
Nucl.Phys.B234:269,1984. }

\lref\nicolai{\paper {H.~Nicolai}{New Linear Systems for 2D Poincar\'e
Supergravities}{\NPB{414}{1994}{299},}{\hh 9309052}.}



\lref\bg{\paper {L.~Baulieu, B.~Grossman}{Monopoles and Topological Field
Theory}{\PLB{214}{1988}{223}.}{}}

\lref\seibergsix{\paper {N.~Seiberg}{Non-trivial Fixed Points of The
Renormalization Group in Six
 Dimensions}{\PLB{390}{1997}{169}}{\hh 9609161}\semi
\paper {O.J.~Ganor, D.R.~Morrison, N.~Seiberg}{
  Branes, Calabi-Yau Spaces, and Toroidal Compactification of the N=1
  Six-Dimensional $E_8$ Theory}{\NPB{487}{1997}{93}}{\hh 9610251}\semi
\paper {O.~Aharony, M.~Berkooz, N.~Seiberg}{Light-Cone
  Description of (2,0) Superconformal Theories in Six
  Dimensions}{Adv. Theor. Math. Phys. {\bf 2} (1998) 119}{\hh 9712117.}}

\lref\cs{\paper {L.~Baulieu}{Chern-Simons Three-Dimensional and
Yang--Mills-Higgs Two-Dimensional Systems as Four-Dimensional Topological
Quantum Field Theories}{\PLB{232}{1989}{473}.}{}}

\lref\beltrami{\paper {L.~Baulieu, M.~Bellon}{Beltrami Parametrization and
String Theory}{\PLB{196}{1987}{142}}{}\semi
\paper {L.~Baulieu, M.~Bellon, R.~Grimm}{Beltrami Parametrization For
Superstrings}{\PLB{198}{1987}{343}}{}\semi
\paper {R.~Grimm}{Left-Right Decomposition of Two-Dimensional Superspace  
Geometry and Its BRS Structure}{Annals Phys. {\bf 200} (1990) 49.}{}}

\lref\bbg{\paper {L.~Baulieu, M.~Bellon, R.~Grimm}{Left-Right Asymmetric
Conformal Anomalies}{\PLB{228}{1989}{325}.}{}}

\lref\bonora{\paper {G.~Bonelli, L.~Bonora, F.~Nesti}{String Interactions
from Matrix String Theory}{\NPB{538}{1999}{100},}{\hh 9807232}\semi
\paper {G.~Bonelli, L.~Bonora, F.~Nesti, A.~Tomasiello}{Matrix String Theory
and its Moduli Space}{}{\hh 9901093.}}

\lref\corrigan{\paper {E.~Corrigan, C.~Devchand, D.B.~Fairlie,
J.~Nuyts}{First Order Equations for Gauge Fields in Spaces of Dimension
Greater Than Four}{\NPB{214}{452}{1983}.}{}}

\lref\acha{\paper {B.S.~Acharya, M.~O'Loughlin, B.~Spence}{Higher
Dimensional Analogues of Donaldson-Witten Theory}{\NPB{503}{1997}{657},}{\hh
9705138}\semi
\paper {B.S.~Acharya, J.M.~Figueroa-O'Farrill, M.~O'Loughlin,  
B.~Spence}{Euclidean
  D-branes and Higher-Dimensional Gauge   Theory}{\NPB{514}{1998}{583},}{\hh
  9707118.}}

\lref\Witr{\paper{E.~Witten}{Introduction to Cohomological Field   Theories}
{Lectures at Workshop on Topological Methods in Physics (Trieste, Italy, Jun
11-25, 1990), \IJMP{A6}{1991}{2775}.}{}}

\lref\ohta{\paper {L.~Baulieu, N.~Ohta}{Worldsheets with Extended
Supersymmetry} {\PLB{391}{1997}{295},}{\hh 9609207}.}

\lref\gravity{\paper {L.~Baulieu}{Transmutation of Pure 2-D Supergravity
Into Topological 2-D Gravity and Other Conformal Theories}
{\PLB{288}{1992}{59},}{\hh 9206019.}}

\lref\wgravity{\paper {L.~Baulieu, M.~Bellon, R.~Grimm}{Some Remarks on  the
Gauging of the Virasoro and   $w_{1+\infty}$
Algebras}{\PLB{260}{1991}{63}.}{}}

\lref\fourd{\paper {E.~Witten}{Topological Quantum Field  
Theory}{\CMP{117}{1988}{353}}{}\semi
\paper {L.~Baulieu, I.M.~Singer}{Topological Yang--Mills Symmetry}{Nucl.
Phys. Proc. Suppl. {\bf 15B} (1988) 12.}{}}

\lref\topo{\paper {L.~Baulieu}{On the Symmetries of Topological Quantum Field
Theories}{\IJMP{A10}{1995}{4483},}{\hh 9504015}\semi
\paper {R.~Dijkgraaf, G.~Moore}{Balanced Topological Field
Theories}{\CMP{185}{1997}{411},}{\hh 9608169.}}

\lref\wwgravity{\paper {I.~Bakas} {The Large $N$ Limit   of Extended
Conformal Symmetries}{\PLB{228}{1989}{57}.}{}}

\lref\wwwgravity{\paper {C.M.~Hull}{Lectures on $\CW$-Gravity,
$\CW$-Geometry and
$\CW$-Strings}{}{\hh 9302110}, and~references therein.}

\lref\wvgravity{\paper {A.~Bilal, V.~Fock, I.~Kogan}{On the origin of
$W$-algebras}{\NPB{359}{1991}{635}.}{}}

\lref\surprises{\paper {E.~Witten} {Surprises with Topological Field
Theories} {Lectures given at ``Strings 90'', Texas A\&M, 1990,}{Preprint
IASSNS-HEP-90/37.}}

\lref\stringsone{\paper {L.~Baulieu, M.B.~Green, E.~Rabinovici}{A Unifying
Topological Action for Heterotic and  Type II Superstring  Theories}
{\PLB{386}{1996}{91},}{\hh 9606080.}}

\lref\stringsN{\paper {L.~Baulieu, M.B.~Green, E.~Rabinovici}{Superstrings
from   Theories with $N>1$ World Sheet Supersymmetry}
{\NPB{498}{1997}{119},}{\hh 9611136.}}

\lref\bks{\paper {L.~Baulieu, H.~Kanno, I.~Singer}{Special Quantum Field
Theories in Eight and Other Dimensions}{\CMP{194}{1998}{149},}{\hh
9704167}\semi
\paper {L.~Baulieu, H.~Kanno, I.~Singer}{Cohomological Yang--Mills Theory
  in Eight Dimensions}{ Talk given at APCTP Winter School on Dualities in
String Theory (Sokcho, Korea, February 24-28, 1997),} {\hh 9705127.}}

\lref\witdyn{\paper {P.~Townsend}{The eleven dimensional supermembrane
revisited}{\PLB{350}{1995}{184},}{\hh9501068}\semi
\paper{E.~Witten}{String Theory Dynamics in Various Dimensions}
{\NPB{443}{1995}{85},}{\hh 9503124}.}

\lref\bfss{\paper {T.~Banks, W.Fischler, S.H.~Shenker,
L.~Susskind}{$M$-Theory as a Matrix Model~:
A~Conjecture}{\PRD{55}{1997}{5112},} {\hh9610043.}}

\lref\seiberg{\paper {N.~Seiberg}{Why is the Matrix Model
Correct?}{\PRL{79}{1997}{3577},} {\hh 9710009.}}

\lref\sen{\paper {A.~Sen}{$D0$ Branes on $T^n$ and Matrix Theory}{Adv.
Theor. Math. Phys.~{\bf 2} (1998) 51,} {\hh 9709220.}}

\lref\laroche{\paper {L.~Baulieu, C.~Laroche} {On Generalized Self-Duality
Equations Towards Supersymmetric   Quantum Field Theories Of
Forms}{\MPL{A13}{1998}{1115},}{\hh  9801014.}}

\lref\bsv{\paper {M.~Bershadsky, V.~Sadov, C.~Vafa} {$D$-Branes and
Topological Field Theories}{\NPB{463} {1996}{420},}{\hh9511222.}}

\lref\vafapuzz{\paper {C.~Vafa}{Puzzles at Large N}{}{\hh 9804172.}}

\lref\dvv{\paper {R.~Dijkgraaf, E.~Verlinde, H.~Verlinde} {Matrix String
Theory}{\NPB{500}{1997}{43},} {\hh9703030.}}

\lref\wynter{\paper {T.~Wynter}{Gauge Fields and Interactions in Matrix
String Theory}{\PLB{415}{1997}{349},}{\hh9709029.}}

\lref\kvh{\paper {I.~Kostov, P.~Vanhove}{Matrix String Partition  
Functions}{}{\hh9809130.}}

\lref\ikkt{\paper {N.~Ishibashi, H.~Kawai, Y.~Kitazawa, A.~Tsuchiya} {A
Large $N$ Reduced Model as Superstring}{\NPB{498} {1997}{467},}{\hh
9612115.}}

\lref\ss{\paper {S.~Sethi, M.~Stern} {$D$-Brane Bound States
Redux}{\CMP{194}{1998} {675},}{\hh 9705046.}}

\lref\mns{\paper {G.~Moore, N.~Nekrasov, S.~Shatashvili} {$D$-particle Bound
States and Generalized Instantons}{} {\hh 9803265.}}

\lref\bsh{\paper {L.~Baulieu, S.~Shatashvili} {Duality from Topological
Symmetry}{} {\hh 9811198.}}

\lref\pawu{ {G.~Parisi, Y.S.~Wu} {}{ Sci. Sinica  {\bf 24} {(1981)} {484}.}}

\lref\coulomb{ {L.~Baulieu, D.~Zwanziger, }   {\it Renormalizable Non-Covariant
Gauges and Coulomb Gauge Limit}, {Nucl.Phys. B {\bf 548 } (1999) 527,} {\hh
9807024}.}

\lref\dan{ {D.~Zwanziger},  {}{Nucl. Phys. B {\bf   139}, (1978) {1}.}{}}

\lref\danzinn{  {J.~Zinn-Justin, D.~Zwanziger, } {}{Nucl. Phys. B  {\bf
295} (1988) {297}.}{}}

\lref\danlau{ {L.~Baulieu, D.~Zwanziger, } {\it Equivalence of Stochastic 
Quantization and the Faddeev-Popov Ansatz,
  }{Nucl. Phys. B  {\bf 193 } (1981) {163}.}{}}

\lref\ukawa{ {A.~Ukawa and M.~Fukugita, } {\it Langevin Simulations Including
Dynamical Quark Loops, }{Phys. Rev. Lett.  {\bf 55 } (1985) {1854}.}{}}

\lref\wilson{ {G.~G.~Batrouni, G.~R.~Katz, A.~S.~Kronfeld, G.~P.~Lepage,
B.~Svetitsky, and K.~G.~Wilson, } {\it Langevin Simulations of Lattice Field
Theories, }{Phys. Rev. {\bf D 32 } (1985) {2736}.}{}}

\lref\munoz{ { A.~Munoz Sudupe, R. F. Alvarez-Estrada, } {}
Phys. Lett. {\bf 164B} (1985) 102; {} {\bf 166B} (1986) 186. }

\lref\okano{ { K.~Okano, } {}
Nucl. Phys. {\bf B289} (1987) 109; {} Prog. Theor. Phys.
suppl. {\bf 111} (1993) 203. }

\lref\singer{
 I.M. Singer, { Comm. of Math. Phys. {\bf 60} (1978) 7.}}

\lref\neu{ {H.~Neuberger,} {Phys. Lett. B {\bf 295}
(1987) {337}.}{}}

\lref\testa{ {M.~Testa,} {}{Phys. Lett. B {\bf 429}
(1998) {349}.}{}}

\lref\Martin{ L.~Baulieu and M. Schaden, {\it Gauge Group TQFT and Improved
Perturbative Yang--Mills Theory}, {  Int. Jour. Mod.  Phys. A {\bf  13}
(1998) 985},   hep-th/9601039.}

\lref\baugros{ {L.~Baulieu, B.~Grossman, } {\it A topological Interpretation
of  Stochastic Quantization} {Phys. Lett. B {\bf  212} {(1988)} {351}.}}

\lref\bautop{ {L.~Baulieu}{ \it Stochastic and Topological Field Theories},
{Phys. Lett. B {\bf   232} (1989) {479}}{}; {}{ \it Topological Field Theories
And Gauge Invariance in Stochastic Quantization}, {Int. Jour. Mod.  Phys. A
{\bf  6} (1991) {2793}.}{}}

\lref\samson{ {L.~Baulieu, S.L.~Shatashvili, { \it Duality from Topological
Symmetry}, {JHEP {\bf 9903} (1999) 011, hep-th/9811198.}}}{}

\lref\halpern{ {H.S.~Chan, M.B.~Halpern}{}, {Phys. Rev. D {\bf   33} (1986)
{540}.}}

\lref\yue{ {Yue-Yu}, {Phys. Rev. D {\bf   33} (1989) {540}.}}

\lref\neuberger{ {H.~Neuberger,} {\it Non-perturbative gauge Invariance},
{ Phys. Lett. B {\bf 175} (1986) {69}.}{}}

\lref\gribov{  {V.N.~Gribov,} {}{Nucl. Phys. B {\bf 139} (1978) {1}.}{}}

\lref\huffel{ {P.H.~Daamgard, H.~Huffel},  {}{Phys. Rep. {\bf 152} (1987)
{227}.}{}}

\lref\damhuff{ {P.H.~Daamgard, H.~Huffel, Eds.},  {Stochastic
Quantization, } World Scientific (1988).}

\lref\namok{  {M.~Namiki and K. Okano, Eds,} {}{Prog. Theor. Phys. Suppl {\bf
111} (1993). {}}{}}

\lref\damtsok{  {P.H.~Damgaard and K. Tsokos,} {}{Nucl. Phys. B {\bf 235}
(1984) {75}.}{}}

\lref\creutz{ {M.~Creutz},  {\it Quarks, Gluons and  Lattices, }  Cambridge
University Press 1983, pp 101-107.}

\lref\zinn{ {J. ~Zinn-Justin, }  {Nucl. Phys. B {\bf  275} (1986) {135}.}}

\lref\shamir{  {Y.~Shamir,  } {\it Lattice Chiral Fermions
  }{ Nucl.  Phys.  Proc.  Suppl.  {\bf } 47 (1996) 212,  hep-lat/9509023; 
V.~Furman, Y.~Shamir, Nucl.Phys. B {\bf 439 } (1995), hep-lat/9405004.}}

 \lref\kaplan{ {D.B.~Kaplan, }  {\it A Method for Simulating Chiral
Fermions on the Lattice,} Phys. Lett. B {\bf 288} (1992) 342; {\it Chiral
Fermions on the Lattice,}  {  Nucl. Phys. B, Proc. Suppl.  {\bf 30} (1993)
597.}}

\lref\neubergerr{ {H.~Neuberger, } {\it Chirality on the Lattice},
hep-lat/9808036.}

 

\Title{ \vbox{\baselineskip12pt\hbox{hep-th/9909006}
\hbox{CERN-TH-99-145}
\hbox{RU-99-486}
\hbox{LPTHE-99-66}
\hbox{NYU-TH-99-8-29}}} {\vbox{\centerline{ QCD$_4 $ From A Five-Dimensional
Point of View    }}}
\centerline{{\bf Laurent Baulieu}$^{\star \dag  \S    }$ and  {\bf  Daniel
Zwanziger}$^{ \ddag}$}
\centerline{baulieu@lpthe.jussieu.fr, Daniel.Zwanziger@nyu.edu}
\vskip 0.5cm
\centerline{\it $^{\star}$LPTHE, Universit{\'e}s P. \& M. Curie (Paris~VI) et
D. Diderot (Paris~VII), Paris,  France,} 


\centerline{\it $^{\dag}$ TH-Division, CERN, 1211 Gen{\`e}ve 23, Switzerland}
\centerline{\it $^{\S}$Department   of  Physics and  Astrophysics, Rutgers 
University,  NY 09955-0849, USA}
\centerline{\it $^{\ddag}$   Physics Department, New York University,
New-York,  NY 10003,  USA}

\medskip
\vskip  1cm
\noindent We propose a 5-dimensional   definition for
the physical $4D$-Yang--Mills theory. 
The
 fifth dimension corresponds to the Monte--Carlo time of numerical
simulations of  QCD$_4 $. 
  The 5-dimensional  theory is     a well-defined topological quantum 
field theory  that  can be renormalized at
 any given finite order of perturbation theory.  The relation to
non-perturbative physics is obtained by
 expressing the theory on a lattice, a la Wilson.   The new    fields  that
must be introduced in the context of a  topological Yang--Mills theory have
a simple lattice expression.  We present a 5-dimensional critical limit for
physical
 correlation functions and  for dynamical  auto-correlations,
which allows new Monte--Carlo algorithm based on the time-step in lattice
units given by
$\e = g_0^{-13/11}$ in pure gluodynamics. The gauge-fixing in five dimensions
is such that no
 Gribov ambiguity occurs. The weight is strictly positive, because all
 ghost fields have parabolic propagators and yield trivial determinants.
 We  indicate how  our  5-dimensional  description      of the     Yang--Mills
 theory  may be extended to  fermions.

\Date{\ }

\def\e{\epsilon}
\def\demi{{1\over 2}}
\def\quart{{1\over 4}}
\def\pa{\partial}
\def\a{\alpha}
\def\b{\beta}
\def\d{\delta}
\def\c{\gamma}
\def\m{\mu}
\def\n{\nu}
\def\r{\rho}
\def\s{\sigma}
\def\l{\lambda}
\def\L{\Lambda}

\def\P{\Psi}
\def\F{\Phi}

\def\w{\wedge}

\newsec{Introduction}

The 4-dimensional Yang--Mills theory seems to suffer  from  logical
contradictions.  In the continuum  formulation,
 one has a gauge-fixed BRST invariant path integral, but one has 
 the famous Gribov ambiguity for large gauge field configurations
\gribov\singer.  One often  discards this problem, since the idea of defining
the theory  as a path integral of a gauge field can only be seriously
advocated in  perturbation theory or for  semi-classical approximations. In
the  lattice formulation,
 which is by construction valid non-pertubatively,
 one chooses as variables the gauge group elements, but  one has yet
another contradiction.  The way the continuum
theory is approached is unclear, and if one tries to do a local
gauge-fixing, the partition function vanishes \neuberger.
 This question is   also  often discarded, since   for
computing  gauge-invariant quantities,  one can factorize the volume of the
gauge group that   is finite on the  lattice.

It is however frustrating not to have continuum and  lattice formulations
which would separately  
 define both gauge-invariant and gauge non-invariant sectors, with a BRST
symmetry  controlling the correspondence of the theory to a physical sector,
and with a natural limit from the lattice to the continuum formulation.

 In this article,  we   will show that, in order to reconcile the continuum and
lattice approaches, it is useful to define the theory in a 5-dimensional
 space, such that the 4-dimensional physical theory lives in a slice of
this extended space. The theory that we will consider passes  the tests that
are obviously needed: in the continuous formulation  it is perturbatively 
renormalizable by power counting without loosing its physical character,
(due to Ward identities),  and the gauge-fixing is no longer subject to the 
Gribov ambiguity; in the lattice formulation,  all fields that one considers 
in  the continuum formulation take their place, and one now  obtains   a 
formulation with a consistent gauge fixing.

The fifth-dimension will be the stochastic time that  Parisi and Wu proposed
long time ago 
 for stochastically quantizing  the Yang--Mills theory through a Langevin
equation \pawu\huffel. For reviews of stochastic quantization, see \damhuff
\namok.  The  5-dimensional theory will
 be a supersymmetric theory of the  topological type, that we can express as
a path integral over 5-dimensional  gauge fields for the continuum
perturbative formulation, or as a lattice gauge theory which now depends on
fields of a topological field theory.  In one of the lattice formulations
that we will present, the stochastic time is discretized.  In the
5-dimensional critical limit controlled by $g_0 \to 0$, both the physical
Euclidean correlation lengths and the dynamical Monte Carlo
auto-correlation times diverge simultaneously.  This allows new algorithms.

 The link   between    topological field theory and   stochastic
quantization  was first
observed in \baugros.
  The general idea is that the $D+1$-dimensional supersymmetric formulation
of a
$D$-dimensional quantum field theory  that is   obtained by expressing the
field quantization by a Langevin equation,  involves the fields of a
topological field
theory in $D+1$ dimensions.  Then, supersymmetric cancellations wash away the
detail of the theory in the bulk, while the relevant aspects of the physical
theory are retained in some  boundary of the
$D+1$-dimensional space. 

Moreover an extra gauge field component is needed to  enforce the
5-dimensional gauge symmetry in the stochastic
  framework   \bautop.  The latter can
be identified as a potential for a drift force along the gauge orbits of the
original $D$-dimensional gauge
  theory, as independently observed in \halpern\bautop\yue. A
gauge-fixing drift force for stochastic quantization was actually    
introduced originally in \dan, as a function of
   the 4-dimensional gauge fields $A_\mu$. But the point is that,
  by promoting the potential for this force to a 5th field component,
and by
  functionally integrating over all its possible values, \bautop, not only
does one not alter  the physical physical sector of the theory, but one
  softens the gauge condition of the $D$-dimensional quantum field
  theory.  This actually gives the desired result that the Gribov ambiguity
  losses is relevance as an obstacle for the gauge fixing of the
  theory.

By postulating that QCD$_4$ should be considered from a 5-dimensional point
of view, we
 will emphasize its  relationship to a topological field theory which 
establishes a pleasant  geometrical framework, both in the continuum and
lattice descriptions. 

Renormalizability of stochastically quantized  scalar $\phi^4$ theory was
demonstrated in  \zinn\ using the BRST operator which encodes the
supersymmetry of the stochastic process.  Renormalizability of the
5-dimensional formulation of gauge theory was demonstrated in  \danzinn\
using the BRST operator for gauge invariance supplemented by graphical
arguments.  In the present article we demonstrate renormalizability using
the complete BRST operator that encodes both the supersymmetry of the
stochastic process and the gauge symmetry of the theory, consistent with
parabolic propagation of all ghost fields.

 The disappearance of the Gribov ambiguity comes as an immediate
consequence of being in five dimensions, in the context of renormalizable
gauges (adapted to the power counting of this dimension). We will check
this result by the verification that the ghost propagators cannot have zero
modes. Alternatively,  the mathematically oriented reader will notice that
the argument of Singer \singer\ for having a Gribov copy phenomenon just
disapears because  the  stochastic processes is defined on an  infinite
interval, even if we have a compact 4-dimensional physical space.

  The
 consistency of our approach is ensured by the topological invariance,
 defined modulo ordinary gauge symmetry. We will also indicate that
 instantons in four dimensions are  replaced by 5-dimensional solitons,
but we leave open the question of anomalies, which is presumably connected
to interesting 5-dimensional topological questions.

 To make contact to non-perturbative physics, we will   formulate the fields
and the symmetry in a 5-dimensional lattice formulation. Not only this
offers a new point of view for the topological symmetries, but this  allows
us to obtain a   discrete lattice formulation with   a BRST invariant 
gauge-fixing, offering thereby a new understanding of the continuous limit.
It also gives a concrete definition of non-perturbative physics.

	In secs. 2-5 we present the continuum 5-dimensional topological formulation. 
In sec. 6 we describe the lattice stochastic formulation including the lattice
Fokker-Planck and Langevin equations, in which the 5th time $t = x_5$ is
identified with the number of Monte Carlo sweeps.  We describe a new Monte
Carlo algorithm based on the 5-dimensional critical limit $g_0 \to 0$ which
controls both physical correlation lengths and dynamical auto-correlation
times, with time-step in lattice units given by $\e \sim g_0^{-13/11}$ in
gluodynamics.  We conjecture that the physical spectrum may be given by the
eigenvalues of the Fokker-Planck hamiltonian.  In sec. 7 we present a
5-dimensional topological lattice formulation.  In sec. 8 we indicate
how our approach extends to fermions, and we conclude with some speculation
on future developments.  Readers who are primarily interested in Monte Carlo
calculations may read sec. 6 independently.

\newsec{The 5-dimensional continuum action}

\subsec{The Langevin equation}

The Langevin equation proposed by Parisi and Wu was \pawu:
\eqn\pw{\eqalign{  \pa_5 A_\m=  D_\l F_{\l \m} + \eta_\m, }} where $
\eta_\m $ is
a Gaussian white noise.  In the stochastic approach, the correlation
functions are  functionals of  $A_\m(x^\m,x^5)$ that are the solutions of
this equation. The 4-dimensional Green functions are obtained  in the limit
where all arguments in the stochastic time are equal, and  $x^5\to \infty$.
Since we
consider a stochastic process, we can as well take initial conditions at
$x^5\sim - \infty$, and define
 the Green functions at any given fixed time, for instance
$x^5=0$. The convergence toward an equilibrium distribution relies on  
ergodicity theorems that accord with physical intuition for which the
relaxation of a gas to the state of maximal entropy is obvious, whatever
of the initial distribution of its constituents.

A difficulty with this  Langevin equation  is that it provides no
restoring force along the gauge orbits, because the Euclidean action $S(A)$
is gauge-invariant.  Consequently the probability escapes to infinity along
the gauge orbits, and there is no normalizable equilibrium probability
distribution. This
may be remedied by modifying \pw\ by the introduction of a gauge-fixing
term that is tangent to the gauge orbit
\dan\ ,
 \eqn\ddd{\eqalign{
\pa_5 A_\m = D_\l F_{\l\m} + D_\m v[A] + \eta_\m.
}} 
The gauge-fixing term $D_\m v[A]$ has the form of an infinitesimal
gauge transformation, and consequently it has no effect on the
expectation-value of any gauge-invariant observable. The infinitesimal
generator $v[A](x,t)$ may, in principle, be
completely arbitrary, and the issue of a ``correct" gauge fixing, without
Gribov copies does not arise.

The last equation may be written in a way which respects gauge invariance
in five dimensions. The idea is to identify $v[A]$ with an independent 5th
field component
\eqn\ddda{\eqalign{
v = A_5(x,t)
}} 
which eventually be fixed in the functional integral by a gauge-fixing term in
the action. Then  the gauge-fixed Langevin equation
\ddd\ may be written as:
\eqn\toppp{\eqalign{    
F_{5\m} - D_\l F_{\l\m} = \eta_\m
}} 
where $F_{5\m} = \pa_5A_\m - D_\m A_5$.  The relation   
\eqn\topa{\eqalign{ 
A_5 = a^{-1}\pa_\l A_\l \ .
}}
will be  imposed in the context of topological field theory  by a
5-dimensional gauge condition.

We would like to obtain a functional integral representation for the
Langevin equation \toppp. This is a non-trivial problem  because there is a
gauge
invariance, and the determinant of the map which connects the $A$'s and the
noise $\eta$'s has   longitudinal zero modes. Instead of attempting to solve
these questions by manipulations on the Langevin equation, we will pass
directly to a 5-dimensional topological quantum field theory. 
  Later with lattice regularization we shall verify the consistency of the
5-dimensional formulation, including the absence of Gribov copies,
and we shall shall show that after elimination of auxiliary fields present in
this theory, one recovers the Langevin equation and the equivalent Fokker-Planck
equation.

 \subsec{The 5-dimensional BRST symmetry}

The expression of the complete symmetry of the theory boils down to the
knowledge of a BRST operator which encodes at once the supersymmetry of the
stochastic process and the gauge symmetry of the theory. This operator is
that of the topological symmetry of a gauge field in five dimensions,
defined modulo   gauge transformations. There is not much choice for
exhibiting such a BRST operator, and the solution was given in \bautop. It
  involves  a fifth dimensional component for the gauge field, that
we denote as $A_5$. Unavoidably,  the vector fermion that is needed to
express the Jacobian of the  constraint \pw\ must be enlarged into a five
dimensional vector $(\P_\m,\P_5)$. If we denote by
$c$ the ordinary Faddeev--Popov ghost and $\F$ its ghost of ghost, the
combined BRST operator for the stochastic supersymmetry and the gauge
symmetry follows therefore
 from the following geometrical equation \bautop:
  \eqn\gfc{\eqalign{ (s+d)(A+c)+\demi[A+c,A+c]&=F+\P_\m dx^\m+\P_ 5
 dx^ 5+\F
\cr
 (s+d)(F+\P_\m dx^\m+\P_ 5 dx^ 5+\F) &=- [A+c,F+\P_\m dx^\m+\P_ 5 dx^ 5+\F]
}} The second equation is the Bianchi identity of the first, thus one has
$s^2=0$ on all fields which ensures the consistency of this symmetry which
mixes the Yang--Mills symmetry with the supersymmetry of the stochastic
process.  Here $c, \P_\m$ and $\P_5$ are fermi ghost fields with ghost
number $N_g = 1$ and $\F$ is a bose ghost field with ghost number $N_g =
2$.
This equation gives, after expansion in ghost number:
\eqn\gfcm{\eqalign{   sA _{ \m } &=\P _{\m } + D _{ \m }c
\cr sA _{ 5 } &=\P _{5 } + D _{ 5 }c
\cr s \P _{\m } &= -D _{ \m }\F -[c,\P_\m] \cr s \P _{5 } &= -D _{ 5 }\F
-[c,\P_5] \cr sc &=
\F-\demi[c,c]  \cr s\F &= - [c ,\F] \ . }} 
It is  identical to the
topological BRST symmetry for a Yang--Mills field in five dimensions.

We need anti-ghosts and lagrange multipliers, and their
$s$-transformations.  They are:
\eqn\ggfc{\eqalign{   s\bar \P _{ \m } &=b _{\m } - [c,\bar\P_\m]
\cr s\bar \P _{ 5 } &=b _{5 } - [c,\bar \P_5]
\cr s b _{ \m } &= -[c,b_\m]+[\F,\bar \P _{ \m }]
\cr s b _{ 5 } &= -[c,b_5]+[\F,\bar \P _{ 5 }]
\cr s \bar\F &= \bar\eta -[c,\bar\F] \cr s \bar\eta &= -[c,\bar\eta]
+[\F,\bar\F] \ . }}  
All fields in \gfcm\ and \ggfc\ are valued in the same
Lie algebra representation as $A$.  In the lattice formulation of this
symmetry, it will be advantageous to redefine the anti-ghost and lagrange
multiplier fields to eliminate the dependance on $c$ in \ggfc, that is,
$s\bar\P=b', \ sb'=0$,
$s\bar \P=b', \ sb'=0$, $s \bar \F=\bar \eta'$, and $s\bar \eta'=0$.

There are five degrees of freedom for the choice of a dynamics, one for
each component
of $A$, with Lagrange multipliers fields $b_\m$ and $b_5$. One of them will
serve to gauge-fix the  ordinary 5-dimensional
gauge invariance, with the condition \topa;   the other four    will 
enforce the Langevin equation \toppp, following the now  standard methods  
of TQFTs.  
The gauge
fixing of the longitudinal modes in $\P$ will use $\bar\eta$ as a Lagrange
multiplier.

\subsec{The action}
The complete  action must be $s$-invariant. We will gauge fix the 5-dimensional 
invariant:
\eqn\top{\eqalign{  
 \int dx^\m dx^5\ \Tr ( F_{5\m} D_\n F^{\m\n})}}
 This term is invariant under any local shift in $A_\mu$ and $A_5$,  since the
Lagrangian can locally be written as $\Tr \pa_\m (F_{5 } ^\nu F_{\n } ^\m)
+\demi\pa_5( F_{\m } ^\nu F_{\n } ^\m )$. Of
course this means that we use special boundary conditions for the
variations, and eventually on the 
$\Psi$'s. The choice of this term may appear as  quite
  intuitive: it  enforces  the idea that the stochastic time is
unobservable,   since it is independent of the metric component $g_{55}$, and it is compatible with Yang--Mills invariance. 
We will use an $s-$exact term for fixing all the invariances and
introducing all relevant 
drift forces of stochastic quantization.    Remarkably, the topological
BRST symmetry just described is precisely
what is needed to do so, that is, to represent a gauge theory in stochastic
quantization by a
functional integral with a high degree of symmetry~\bautop.


To impose all relevant gauge conditions,
 we take the
 BRST-exact action:
\eqn\tgf{\eqalign{ I= 
 \int dx^\m dx^5\ s\ \Tr \big( &
\bar \P^\m   ( F_{5\m} - D_\l F_{\l\m} -\demi b_\m)\cr& +\bar \F    (a'\P_5-
D_\l \P _\l + {{\b}\over 2} [\bar\eta,\F])\cr& +\bar \P^5 ( a A_{5 }-
 \pa_\l A_\l )\big) . }} 
The consistency of this  choice of this gauge function will be further
justified by showing that it gives a  perturbatively renormalizable theory.
 Here $a$, $a'$  and $\b$ are arbitrary parameters.  When $\b$ is non-zero,
quartic interactions are introduced.  Of course $\b=0$ gives simpler 
expressions,
 but nothing forbids $\b\neq 0$, ($\b=0$ might be a stable fixed point under
renormalization). [Alternatively one may make a linear gauge choice by setting
$\b = 0$ and taking $\p_\l \P _\l$ instead of $D_\l \P _\l$.]  The first two
terms  are invariant
 under 5-dimensional gauge transformations. The first term concentrates
the path integral around the solutions of \toppp,  modulo ordinary gauge
transformations, while the  second term fixes
 in a gauge-covariant way the internal gauge invariance of $\P_\m$
 that one detects in the BRST variation $s\P_\m=D_\m\F+...$. The third
 term fixes the gauge invariance for $A_\m,A_5$.  As we will see in section
3,  all these
terms are essentially determined by symmetry   and   power counting
requirements.

  We may introduce a background Yang--Mills symmetry, which transform all
fields, but $A$, in the adjoint representation, while $A$ transform as a
gauge field. The shortest way to represent such a symmetry, is to define it
through a background BRST symmetry, with generator $\s$ and  background 
ghost $\omega$ (which do not appear in the action). Since $s$ and $\s$ must
anticommute, and
$\s^2=0$, we easily find that we obtain both the action of  $s$ and $\s$ by
extending \gfc\ into:
 \eqn\gfcs{\eqalign{ &(s+\s+d)(A+c+\omega)+\demi[A+c+\omega,A+c+\omega] =F+\P_\m
dx^\m+\P_ 5
 dx^ 5+\F
\cr
 &(s+\s+d)(F+\P_\m dx^\m+\P_ 5 dx^ 5+\F) =- [A+c+\omega,F+\P_\m dx^\m+\P_ 5 dx^
5+\F] }} When one expands the latter equation one must   assign a new ghost
number to $\s$ and $\omega$, which is independent of that carried by  $s$
and the
propagating fields. This determines all transformations   of  $s$ and
$\s$ on all fields, with $s^2=\s^2=s\s +\s s=0$. 
One then observes that the first two terms in \tgf\  are not only
$s$-invariant as $s$-exact terms, but they are also $\s$-invariant, without
being $\s$-exact. These terms involve 
gauge-covariant  gauge functions for $A$ and $\P$ in  five dimensions.  The
aim of the renormalization proof, using power counting, will be to show that
these terms, up to multiplicative rescalings, are the only ones which
satisfy both $s$ and $\s$-invariances. The third term is not $\s$-invariant,
and it will be necessary to prove that when the parameter $a$ changes, this
does not affect the sector of the theory defined by 
 $s$ and $\s$ invariances.

To investigate the properties of our action, we  must expand \tgf, which is
a simple exercise. One gets:
 \eqn\tgfB{\eqalign{  I = 
 \int dx^\m dx^5\ \Tr \big(& - \demi b_\m b_\m+ b_\m ( F_{5\m} - D_\l
F_{\l\m} )\cr & +\bar \F ( - a'D_5 \F + D_\m D_\m \F-[\P_\m,\P_\m]) -
\demi \F [\bar \P_\m,\bar \P_\m]+{{\b}\over 2}[\F,\bar\F][\F,\bar\F]\cr&
+(b_5-[c,\bar \P_5]) ( a A_{5 } - \pa_\m A_\m )\cr & -\bar \P_\m ( D_{[5}
\P_{\m]} - D_\l D_{[\l } \P_{\m]} - [\P_\l, F_{\l\m}])
\cr&
 +\bar \eta (a'\P_5 - D_\m \P _\m +{{\b}\over 2} [\bar\eta,\F])\cr& -\bar
\P^5 ( a (\P_{5 }+D_5 c) - \pa_\m \P_\m - \pa_\m D_\m c)\big). }}

        In order to see the dynamical content of this action, we identify
and eliminate the auxiliary fields.  The equation of motion of $b_5$ gives
back the gauge-fixing condition that we wish to impose:
\eqn\drift{\eqalign{   a A_{5 }  =  \pa_\l A_\l  . }} Moreover the equation
of motion of $b_\m$ has the same form as the Langevin equation,
\eqn\tst{\eqalign{  F_{5\m} - D_\l F_{\l\m} = b_\m .}} We   also eliminate
$\P_5$ and  $\bar \eta$ and we get,
\eqn\tgfaa{\eqalign{
 a'\P_5 = D_\m \P_\m - {{\b}\over 2} [\bar\eta,\F] }} 

\eqn\tgfuu{\eqalign{ 
a'\bar \eta   =a\bar \P_5+D_\m\bar \P_\m.}} Thus    
$A_5$ can be expressed as a function of $A_\m$, and $\P_5$ can be  
expressed as a function of $\P_\m$.  This is the key of a gauge-fixing that
does not suffer from the Gribov ambiguity, and  the gauge condition  $ a A_{5
}  =  \pa_\l A_\l$ combines the virtues of axial and Laudau gauges. By
integrating over all values of $A_5$,   the Faddeev--Popov zero modes that
one encounters in the
genuine 4-dimensional theory just disappear. It is tempting to compare the 
integration over $A_5$ to the integration over moduli  that also solves 
the problem  of   zero modes of reparametrization ghosts in string theory. 

  We will give much more details about the elimination of
Gribov copies in sec. 7,
although a Fourier transformation on the variable $x^5$ could help
understanding directly how the multiple intersections of gauge orbit that
occur in four dimensions could be split on different orbits in five
dimensions. 

 Finally, after integrating out  $b_\m$ and $b_5$,   we get:
\eqn\tgfff{\eqalign{  I +\int dx^\m dx^5\ \Tr ( F_{5\m}     D_\l F^{\l\m})
\sim &
 \int dx^\m dx^5\ \Tr \big(-\demi F_{5 _\m} F^{5 _\m} -\demi D_\n
 F^{\n }_\m D_\r F^{\r\m } 
\cr &
 +\bar \F ( a'D_5 \F+ D_\m D^\m
 \F)
\cr & +\bar \F [\P_\m,\P^\m]) +\demi \F [\bar \P_\m,\bar \P^\m]
 +{{\b}\over 2}[\F,\bar\F][\F,\bar\F] 
\cr &
 -\bar \P^\m ( D_{ 5} \P_{\m
 } + D^\n D_{[\m } \P_{\n]}+[\P^\n, F_{\m\n}])
\cr & +a'^{-1}D_\m\bar\P^\m ( D_\n\P ^{\n }-{{\b}\over 2} [\bar\P_5,\F])
\cr &
 -\bar \P^5 ( a D_5 c +\pa^\m D_\m c+[A_\m,\P_\m])\ \big) }} One has 
$A_5=a^{-1} \pa_\l A_\l$ in this action.

        To derive Feynman rules, we examine the part of the resulting action
that is quadratic in the fields.  It is given by
\eqn\tste{\eqalign{ I_0 = I_A + I_\P + I_\F + I_c}}
\eqn\tstee{\eqalign{ I_A =& \int dx^\m dx^5
\Tr \big( \demi   (\pa_5 A_\m)^2 + 
 \demi (\pa_\l \pa_\l A_\m )^2 + \demi (a^{-2} -1) (\pa_\m \pa_\l A_\l )^2
\big)\cr I_\P =& \int dx^\m dx^5 \Tr \big(
\bar \P_\m   ( - \pa_5 \P_\m + \pa_\l \pa _\l \P_\m   + (a'^{-1} -1)\pa_\m
\pa_\l \P_\l) \big) \cr I_\F =& \int dx^\m dx^5  
\Tr \big( \bar \F    ( - a'\pa_5 \F+ \pa_\m \pa_\m \F) \big) \cr I_c =& \int
dx^\m dx^5 \Tr \bar \P^5\big ( - a \pa_5 c + \pa_\m \pa_\m
 c\big)}}
 where we have dropped exact derivatives in $I_A$.

        An important observation is that all ghost-antighost pairs,
$(\P_\m , \bar\P_\m)$, ($\F , \bar\F$) and ($c  ,\bar\P_5$)  have a free action
which is parabolic: it is first order in $\pa_5$, and the corresponding
matrix of spatial derivatives is a negative operator. Consequently all ghost
propagators are retarded, $D(x, t) = 0$ for $ t = x_5 < 0$. For example for
$a = 1$, all ghost-antighost pairs have the momentum-space propagator
$(i\omega + k^2)^{-1}$, with fourier transform $D(x, t) = \theta(t) (4 \pi
t)^{-2} \exp(-x^2/4t) $, where
$\theta(t)$ is the step function.  Moreover ghost number is conserved, and
an arrow which represents the flow of ghost charge may be assigned to each
ghost line: it points from the past to the future.  On the other hand, the
action of the $A$-field is second order in $\pa_5,$ so
$A$ lines freely move forward and backward in time.  (Its momentum-space
propagator is $\Big (\omega^2 + (k^2)^2\Big )^{-1} = (2k^2)^{-1} \Big ((i\omega + k^2)^{-1}
+ (-i\omega + k^2)^{-1}\Big ).$) Ghost number is also conserved at every vertex. 
Consequently in a diagrammatic expansion, every ghost line may be followed
from the point where it enters a diagram from the past, as it moves
monotonically from the past to the future, until it exits into the future. 
(The transformation of a 
$\F$ into two $\P$'s and back is allowed.)  It follows that in a diagram
containing only external $A$-lines, there can be no closed ghost loops apart
from tadples.  With dimensional regularization the tadpole diagrams vanish,
and in this representation of gauge theory, ghost diagrams provide accounting
checks on the renormalization constants that are expressed in Ward identities,
as usual, but they do not appear in the expansion of correlation functions the
$A$-field.  With lattice regularization, the tadpole diagrams yield the
famous Ito term.

Another way to see that the ghost determinants have no Gribov problems is
to note that the initial-value problem  for  a 
parabolic equation has   a unique
solution.  In the section devoted to lattice regularization, we will indeed
integrate out the ghosts exactly, and explicitly  verify the absence of
Gribov copies.

We now turn to the question of understanding the renormalization of this
action.

   \newsec{Perturbative renormalization}

The action \tgfB\ is renormalizable by power counting. Indeed,   the form
of the  propagators implies that the canonical dimension of all fields with
an index $5$ is equal to $2$, while other fields have dimensions $1$. It
follows that we start from an action where the coupling constant has
dimension $0$, and we can apply the general result that it can be
renormalized order by order in perturbation theory by a finite number of
counter-terms. 

The main question is of course to investigate the structure of the
counter-terms. We will only sketch the demonstration that this action is
actually stable under renormalization, which means that the generating
functional of Green functions can be defined, order by order in perturbation
theory, while satisfying the same Ward identities as the
 action \tgf, corresponding  to its invariance under  $s$-symmetry,
 $SO(4) $-symmetry, ghost number conservation,
 and to the choice of a linear condition $aA_5-\partial_\mu A_\mu$ in the
sector which violates the
$\s$-invariance. 

The action \tgfB\  that determines   perturbation theory can indeed be  be
split in two parts:
\eqn\th{\eqalign{ I\sim & \hat
 I(A_\m,A_5,\P_\m,\bar\P_\m,b_\m,\P_5,\F,\bar\F,\bar\eta)\cr &
+\Big (b_5(aA_5-\partial_\mu A_\m)
 -\bar \P^5 ( a \pa_5 c -D^\m \pa_\m c + a\Psi _5 -\pa_\m \P_\m)\Big ). }}
The first
term, where the fields $A_5$ and $\P_5$ must be understood as independent
fields, (before eliminating $b_5$), is $s$-  and
$\s$-invariant, while   the last part is only $s$-invariant. In order to
achieve the algebraic set-up which ensures perturbative renormalizability
while maintaining the form of the action (up to so-called multiplicative
renormalization), it is  enough to remark that one  can write the second
term as:
\eqn\th{\eqalign{ b_5(aA_5-\partial_\mu A_\m)&
 -\bar \P^5 ( a \pa_5 c -D^\m \pa_\m c-a\Psi _5 -\pa_\m \P_\m)\cr=&
b_5(aA_5-\partial_\mu A_\m) 
 -\bar \P^5 ( a D_5  -\pa^\m D_\m )c'}}
 where 
$c'=c+( a \pa_5  -D^\m \pa_\m )^{-1} (a\Psi _5 -\pa_\m \P_\m)$. The change of
variable $c\to c'$ is perturbatively well defined, with a trivial Jacobian,
because the
operator $ a \pa_5  -D^\m \pa_\m$ is parabolic. (This will be explained in
detail in section 7, devoted to the lattice regularization.)

If we now introduce an ordinary BRST symmetry $s'$, with $c'$ as a Faddeev--Popov ghost,  we are almost in the ordinary situation of the
renormalization, except that we have more fields. We have that $s'$ acts on
$A_\m,A_5,\P_\m,\bar\P_\m,b_\m,\P_5,\F,\bar\F,\bar\eta$ as ordinary gauge
transformations with the parameter equal to $c'$, $s'c' = -c'c'$, 
$s'\bar \P_5=b_5$ and $s' b_5 =0$. The action \th\ can be written as: 
\eqn\th{\eqalign{ I\sim & \hat
 I(A_\m,A_5,\P_\m,\bar\P_\m,b_\m,\P_5,\F,\bar\F,\bar\eta)\cr & +s'  \big(
\bar \P^5    (aA_5-\partial_\mu A_\m)\big) }} The first term is $s'$-invariant
as a consequence of its
$\s$-invariance: it   corresponds to the cohomology with ghost number zero
of $s'$, restricted to the terms  with the correct power counting (6 with our
power assignmant), while the second term is $s'$-exact, and will remain stable
order by order in perturbation theory  because of the linear gauge
condition $aA_5-\partial_\mu A_\m$. 
Notice that $\hat I$ depends on $a'$, but $A_\mu$-dependent observables
are independent of this parameter, as will be shown in sec. 7.

The complete proof that the theory can be renormalized multiplicatively  is
just a slight  amplification  of the
standard proof of the renormalizability of the 4-dimensional theory in a
linear gauge. It involves   introducting   sources for all $s'$
variations, as well as using the equation of motion of the antighost
$\bar \P_5$ as a Ward identity. When the Ward identity of the
$s'$-invariance 
 is combined with that of the topological 
 $s$-invariance, using locality, one can   prove from  purely algebraic
considerations
that the counter-terms  can be built order by order in perturbation theory
such that, when they are added to the starting action \tgf, they provide an
identical  action, up to   mere    multiplicative renormalization constants
for the fields and parameters.  Essential is in this proof is the use of a
linear gauge
function $aA_5-\pa_\m A_\m$.

	A notable feature of the renormalization is that besides the
multiplicative renormalization of the fields and coupling constants
there is also a renormalization of the stochastic time
\eqn\rst{\eqalign{
t = Z_t t_r \ .
}} 

\newsec{Observables and gauge invariance}
 
It is now clear how observables should be defined: they are the cohomology
with ghost number zero  of $s'$; moreover, they must be computed at a same
value of $x^5$. Using  the translation invariance under $x^5$, and 
provided that the stochastic process starts at $x^5=-\infty$, the physical 
theory   can be
defined in any given slice of the 5-dimensional manifold.  This points
out the relevance of the boundary term \top, which is independent of the
metric component
$g_{55}$. Perturbatively,  the equal-time gauge-invariant correlation
functions are guaranteed  to agree with those calculated    in 4-dimensional
theory  by the usual Faddeev-Popov method.  On the other hand,   gauge
non-invariant correlation functions  are not expected   to agree within the
framework  of local renormalizable 5-dimensional gauges that we consider in
this paper 
\foot{To recover the local 4-dimensional Faddeev--Popov distribution, one
needs   a non-local 5-dimensional gauge-fixing on $A_5$   \danlau.}.

There are also Green functions that can be computed with fields at different
values of $x^5$, as well as the Green functions that involves the
topological ghosts. We leave their interpretation as an interesting open
question, although they are not directly relevant to physics.

Once the renormalization has been done properly, one
can integrate  out all the ghost fields of the type
$\P$, $\F$ and $c$, provided that one considers observables which depend on
$A_\m$ only. This follows from the parabolic or retarded behavior of all ghost
propagators in the 5-dimensional theory.  It will be demonstrated by explicit
calculation with lattice regularization in section 7 that, when the ghost
fields are integrated out, only the tadpole diagram survives so the ghost
determinent is trivial and contributes only a local term to the effective
action which is in fact the famous Ito term. This result, which explains the
disappearance of the problem of Gribov copies, has an immediate application:
mean values of observables are independent of the parameter $a'$. It should be
clear however that  the ghosts and ghosts of ghosts are  nevertheless
necessary to unveil the topological and gauge properties of the theory and to
control its Ward identities.  

The other question is obviously that of the independence of physical
expectation values of the gauge parameter $a$ which appears in
the gauge function $aA_5-\pa_\m A_\m$. To prove this, one observes
that the observables are  $s'$-invariant without being  $s'$-exact, while
the $a$-dependence is through an $s'$-exact term, so the
standard BRST method based on Ward identities applies. One can also prove
directly the non-renormalization of the $s'$-exact term in \th.

We mention the possibility of introducing  the
interpolating gauges used in \coulomb\ for  defining the ``physical''
Coulomb gauge as the limit of a renormalizable gauge. 
  
Finally, one must of course verify that no anomaly in the $s$- and
$\s$-symmetries can occur. The absence of  $s$-anomalies is quite obvious 
if one examines the consistency conditions for a topological symmetry;  for 
the background Yang--Mills symmetry, no anomaly  is expected,  since we are
in the case of a pure  Yang--Mills theory, without 4-dimensional chiral
fermion. The inclusion of spinors is a most interesting question, on
which we will comment in the last section.

\newsec { Non-perturbative aspects}

In the previous section,  we have verified that perturbation theory is as
well defined in
the 5-dimensional approach as it is in the 4-dimensional one.

\subsec{Global properties of gauge-fixing}

	The gauge choice made here has the global property of a restoring force
because it is derivable from a ``minimizing'' functional.  Consider
the functional 
\eqn\mf{\eqalign{
\CF[A] = (2a)^{-1}(A, A) = (2a)^{-1} \int d^4x A^2,
}}
which is proportional to the Hilbert square norm.  Here $a > 0$ is a gauge
parameter. (More generally, one can take
$\CF[A] = \int d^4x A_\l\a_{\l\m}A_\m $, where $\a$ is a strictly positive
symmetric matrix that characterizes a class of interpolating gauges
\coulomb.)  The gauge condition used here, $aA_5 = \p_\m
A_\m$, may be expressed as
\eqn\mfbb{\eqalign{
A_5(x) = - G(x) \CF,
}}
where
\eqn\mfaa{\eqalign{
G(x) \equiv - D_\m { {\delta } \over {\delta A_\m}(x) }
}}
is the generator of local gauge transformations.  It satisfies the Lie
algebra commutation relations
\eqn\mfaa{\eqalign{
[G^a(x), G^b(y)] = \d(x-y) f^{abc} G^c(x) \ . 
}}
Indeed, one can verify that 
\eqn\mfbz{\eqalign{
G(x) \CF = - a^{-1}\p_\m A_\m(x) .
}}

To see  that the gauge-fixing force is globally restoring,    
consider the flow defined by the gauge-fixing force alone
 \eqn\dddaa{\eqalign{
\p_5 A_\m = D_\m v = - D_\m G \CF ,
}}
in the Langevin equation \ddd.  Under this flow the minimizing functional
$\CF$ decreases monotonically, since
\eqn\dddb{\eqalign{
\pa_5 \CF[A] &= ({ {\delta \CF} \over {\delta A_\m} }, \p_5 A_\m)
= -({ {\delta \CF} \over {\delta A_\m} }, D_\m G \CF) \cr
 &= -  (G\CF, G\CF) \leq 0 .
}} 

	A consequence of our  gauge choice is that the Langevin equation 
\toppp\ is  parabolic.  Indeed, let 
$A_5 = a^{-1} \p_\m A_\m$ be substituted into this equation,  and
consider the highest derivatives on the right-hand side,
\eqn\topp{\eqalign{ {{\pa A_\m }\over { \pa t}} = \pa_\l \pa_\l A_\m +
(a^{-1} - 1)\pa_\m \pa_\l A_\l +\ldots \ ,
}} 
where the dots represent lower order derivatives.  One sees that for
$a > 0$, the operator
that appears on the right is   negative,   which assures convergence at
large $t$.

\subsec{How the Gribov problem is solved}

Let us now  discuss  the Gribov question in more detail, and explain how
our use of the gauge function $aA_5-\pa_\m A_\m$ escapes the ambiguity.

 The origin of this ambiguity in the 4-dimensional formulation, as
given  by
Singer \singer, was the compactness of the (Euclidian) space. It  does not
apply anymore, since the 5-dimensional formulation is necessarily non-compact: 
the system requires an infinite amount of time to relax to equilibrium.  In
our formulation the stochastic time runs from
$-\infty$ to $\infty$, and one evaluates the Green function at an arbitrary
intermediate time, say $x^5=0$.

Then, there is the  explicit argument, which we already mentioned,  of
considering all ghost
propagators, whether they are the $\P$ and $\F$ topological 
ghosts or the Faddeev--Popov ghost: by taking
the suitable boundary conditions at  $x_5 = -\infty$, no zero modes
can occur since we have parabolic propagation. It is a convincing
argument, and we will check it in great detail in section 7, in the
framework of the lattice regularization of our 5-dimensional formulation. We
will prove that the integration on all ghosts does not lead to zero modes that
would make ambiguous their elimination. In our opinion, this a concrete
verification of the assertion that the Gribov problem is solved.

	To see intuitively where the gauge-fixing force concentrates the
probability, observe from   \dddb\ that the gauge fixing force is
in equilibrium only where $G(x)\CF = 0$.  This equilibrium may be
stable or unstable, but the gauge fixing force drives the system toward
stable equilibrium only, namely where the second variation of $\CF[A]$ under
infinitesimal gauge transformation is a positive matrix.  This is equivalent
to the condition that the Faddeev-Popov matrix
\eqn\dddd{\eqalign{
M^{ab}(x,y) = G^a(x) \ G^b(y)  \CF   
 \, }}
has positive eigenvalues. With 
$G\CF = - a^{-1} \p_\m A_\m$, the matrix
$M$ and the condition for stable equilibrium are given by
\eqn\ddde{\eqalign{
M = - a^{-1} \p_\m D_\m \geq 0 .
}}
Thus the conditions for stable equilibrium are (a) transversality of the
vector potential, $\p_\m A_\m = 0$ and (b) positivity of the Faddeev--Popov
operator,
$- \p_\m D_\m \geq 0$.  These two conditions define the Gribov region, a
region which Gribov suggested (not quite correctly) was free of copies.  As
long as the gauge-parameter $a > 0$ is finite, the gauge-fixing is soft in the
sense that the weight on each gauge orbit has a spread that is centered
on the Gribov region.  As
$a$ approaches 0, the equilibrium probability gets concentrated close to the
Gribov region.  The gauge-fixing force drives the system by steepest descent
along a gauge orbit toward a relative or absolute minimum of the functional
$\CI_{[A]}[g] \equiv \CF[{^g A}]$, so Gribov copies that are unstable
equilibria (saddle points of $\CI_{[A]}[g]$) are avoided in the limit $a \to
0$.  However the relative minima are in fact Gribov copies of the absolute
minimum, so in the limit $a \to 0$ this gauge-fixing distributes
the probability in some way among these relative minimum, possibly even
entirely at the absolute minimum.  However the validity of the present
approach in no way depends on taking the limit $a \to 0$.

It is interesting to ask how instantons play a role in the
5-dimensional presentation, for instance in a semi-classical
approximation.
Their interpretation is actually quite simple. 
The bosonic  part of our action is a sum of squares,
namely
$|F_{5\mu}|^2   +  |D_\lambda F_{\lambda\m}|^2   $.
 It  obviously gives and  an absolute minimum when it vanishes. This occurs 
for $x^5$-independent solutions, with $A_5=\pa_\m A_\m=0$ and satisfying 
 $D_\lambda F_{\lambda\m}=0$. This means that the classical 4-dimensional
instanton solutions also minimize the 5-dimensional path integral: they are
the  solitons of the 5-dimensional theory.

 We are therefore in a situation where, from various points of view,  the
path integral seems well defined. We   can now  seriously consider its
evaluation for
``large'' field configurations, since the question of duplicating
orbits does not occur. It is however clear that the space of
connections is not appropriate for defining a meaningful measure, and
we will shortly look at the lattice formulation of this extended
version of the Yang--Mill theory,  keeping in mind that the use of the
fifth time allows a more complete presentation of the
theory.

\newsec{Stochastic lattice gauge theory and new Monte Carlo algorithm}

\subsec{Lattice gauge theory with discrete time}

	Numerical calculations in lattice gauge theory are effected by Monte
Carlo methods which rely on the simulation of a stochastic process.  The
stochastic process in question is specified by a matrix 
$\CT(U_{t+1}, U_t) > 0$ of transition probabilities $U_t \to U_{t+1}$,
on configurations of ``horizontal'' link variables, $U_t = \{U_{x,t,\m}\}$ and
$U_{t+1} = \{U_{x,t+1,\m}\}$, where $\m = 1,...4$.  
Consider a transition probability of the form
\eqn\fdc{\eqalign{
\CT(U_{t+1},U_t) = \CN \exp\{- \b \sum_{x,\m} {\rm Re Tr}   
\Big (I - (U_{x,t+1,\m}^{-1}V_{x,t,\m}) \ \Big )\  \} \ ,
}}
where $\b$ is a positive parameter.  Here $V_{x,t,\m} = V_{x,\m}(U_t)$ is a
group element that depends on the horizontal configuration at time $t$ that
will be specified shortly.  For appropriately chosen normalization constant
$\CN$, this expression satisfies the requirement that the sum of transition
probabilities out of any state is unity 
\eqn\fdca{\eqalign{
\int \prod_{x,\m}dU_{x, t+1, \m }\CT(U_{t+1},U_t) = 1 \ .
}}
This follows from invariance of the Haar measure under
translation on the group, 
$dU_{x, t+1, \m } = d{U'}_{x, t+1, \m }$, where
${U'}_{x, t+1, \m } = V_{x,t,\m}^{-1}U_{x,t+1,\m}$.  

	 We set
\eqn\fdcf{\eqalign{
V_{x,t,\m} =  U_{x,t,5}^{-1} \exp(\e f_{x,t,\m})
U_{x,t,\m}U_{x+\hat{\m},t,5} , 
}}
where $\e$ is another positive parameter, and
 $f_{x,t,\m}$ is a lattice analog of the
continuum drift force $(D_\l F_{\l\m})^{\rm cont}$ that is specified below. 
This gives
\eqn\fdcfa{\eqalign{
\CT(U_{t+1},U_t) = \CN \exp\{- \b \sum_{x,\m} {\rm Re Tr}   
\Big (I - \exp(\e f_{x,t,\m})U_{x,t,\m} U_{x+\hat{\m},t,5} 
       U_{x,t+1,\m}^{-1} U_{x,t,5}^{-1} \Big ) \  \} \ .
}}
where the product of 4 $U$'s is the transporter around the plaquette in the
($\m$-5) plane, starting and ending at the site $(x,t)$.  This expression is
manifestly invariant under 5-dimensional local gauge transformations
provided that $f_{x,t,\m}$ transforms like a site
variable at $x$ in the adjoint representation,
\eqn\dfvg{\eqalign{ 
f_{x,t,\m} \to {^gf_{x,t,\m}} = (g_{x,t})^{-1}f_{x,t,\m} \ g_{x,t} 
\ , }} 
which will be the case.

	In the last expression for $\CT$, the variables $U_{x,t,5}$
associated to time-like links serve merely to effect a gauge
transformation
$g_{x,t} = U_{x,t,5}$ on the variables $U_{x,t+1,\m}$ associated to
the space-like links.  Consequently the $U_{x,t,5}$ may be assigned
arbitrarily without affecting expectation values of the observables which
are the gauge-invariant functions of the space-like link variables at a
fixed time.  We shall shortly gauge-fix $U_{x,t,5}$ by a lattice analog of
continuum gauge fixing implemented above, where by $U_{x,t,5}$ will be
expressed in terms of the $U_{x,t,\m}$ so that eq. \fdca\  is satisfied.

	Recall that the 5-dimensional continuum theory contains no dimensionful
constant, apart from the cut-off, when engineering dimensions are assigned
according to $[t] = [x^2]$.  Consequently the discretized action should
depend on the lattice spacings only through the ratio
\eqn\fdcd{\eqalign{
\e \equiv {{a_t}\over{a_s^2}}  \ ,
}}
which defines the parameter $\e$.  Here $a_s$ and $a_t$ are the lattice
spacings in the space and time directions respectively, with $a_t$ being the
``time'' for a single sweep of the lattice in a Monte Carlo updating.  We write 
\eqn\fdcb{\eqalign{
U_{x,t,\m} & \sim \exp(g_0  a_s  A_{x,t,\m}^{\rm cont})   \cr
U_{x,t,5} & \sim \exp(g_0   a_t   A_{x,t,5}^{\rm cont}) \ ,
}} 
where $g_0$ is the unrenormalized coupling constant, and ``cont"
designates continuum perturbative variables.  

	To verify this point, we estimate the
quantities that appear in eqs. \fdc\ and \fdcf.  The lattice drift force,
specified below, is of order
\eqn\fdch{\eqalign{ 
f_{x,t,\m} & \sim g_0 a_s^3  (D_\l F_{\l\m})^{\rm cont} \ .
}}
This gives
\eqn\fdchaa{\eqalign{ 
\exp(\e f_{x,t,\m}) & \sim 
\exp\Big (g_0 a_s a_t \ (D_\l F_{\l\m})^{\rm cont}\Big ) \cr
U_{x,t,\m} U_{x+\hat{\m},t,5} U_{x,t+1,\m}^{-1} U_{x,t,5}^{-1}
 & \sim \exp\Big ( \ - \ g_0 a_s a_t \ (F_{5,\m})^{\rm cont}\Big ) 
\ , }}
and finally
\eqn\fdcha{\eqalign{ 
{\rm ReTr}\Big (1 - \exp(\e f_{x,t,\m}) \ 
U_{x,t,\m} U_{x+\hat{\m},t,5} U_{x,t+1,\m}^{-1} U_{x,t,5}^{-1}\Big )   \cr
\sim  \Big ({{g_0 a_s a_t}\over {2}}\Big)^2\Big ( (F_{5\m}^a - D_\l F_{\l\m}^a)^{\rm cont}\Big )^2
\ , }}
where a sum on $\m$ and $a$ is understood.  We have  
\eqn\fdchab{\eqalign{ 
\prod_t \CT(U_{t+1}, U_t) = \exp(- \b I) \ ,
}}
where the five-dimensional discretized action is given by
\eqn\nae{\eqalign{
I \equiv \sum_{x,t,\m} {\rm Re \ Tr}
\Big (1 -  \exp(\e f_{x,t,\m})
U_{x,t,\m}U_{x+\hat{\m},t,5} U_{x,t+1,\m}^{-1}U_{x,t,5}^{-1}  \Big ) \ .   
}}
We define
\eqn\fdce{\eqalign{
\b \equiv {{1}\over {g_0^2\e }} = {{a_s^2}\over {g_0^2a_t}} ,
}}
and we obtain
\eqn\naf{\eqalign{
\b I \sim \quart a_t a_s^4 \sum_{x,t,\m} 
\Big ( (F_{5\m}^a - D_\l F_{\l\m}^a)^{\rm cont}\Big )^2 \ ,
}}
which is the correct volume element, $a_t a_s^4$, and the correct
normalization of the 5-dimensional action.

	There remains to specify the lattice drift force $f_{x,\m}$ and the
gauge-fixing of $U_{x,t,5}$.  For this purpose we shall establish a
correspondence between lattice and continuum quantities.  Let $S = S(U)$ be
the gauge-invariant 4-dimensional Euclidean lattice action normalized to
\eqn\dfvha{\eqalign{
S(U) \Longleftrightarrow g_0^2 S^{\rm cont} = { g_0^2 \over {4}}\int d^4x 
(F_{\m\n}^{a \ {\rm cont} })^2 \ ,
}}
which depends only on the horizontal link variables $U = \{U_{x,\m}\}$ at a
fixed time.  It may be the Wilson plaquette action
\eqn\dfvhb{\eqalign{ 
S_{\rm W} = 2 \sum_p {\rm Re \ Tr}(I - U_p)
}}
for pure
gluodynamics, where the sum extends over all plaquettes $p$, or the
effective action that results from integrating out the quark degrees of
freedom,
\eqn\dfvh{\eqalign{
S(U) = S_{\rm W} - \ g_0^2{\rm Tr \ ln} (\g_\m D_\m + m) \ ,
}}
where $\g_\m D_\m + m$ represents the lattice Dirac operator of choice. 
Let the lattice color-``electric'' field operator
$E_{x,\m, a}$ be defined by  
\eqn\eeaa{\eqalign{
E_{x,\m, a}  \equiv
 (t_a U_{x,\m})_{\a \b} { {\p \ \ \ \ \ \ \ \ \ \ } 
\over { \p(U_{x,\m})_{\a \b} } } \ ,
 }}
where $[t_a, t_b] = f^{abc}t_c$, and ${\rm tr}(t_a t_b) = - (1/2)\d_{ab}$.
It has 4 components, $\m = 1,...4$, because our perspective is 5-dimensional.
With 
$(U_{x,\m})_{\a \b} \sim \delta_{\a \b} 
+  (t_a)_{\a \b} \ g_0 a_s A_{x,\m}^{a \ {\rm cont}}$,
we have the correspondence between lattice and continuum quantities,
\eqn\eea{\eqalign{
 E_{x,\m, a} S(U) \Longleftrightarrow
 g_0 a_s^3 \ { {\d S^{\rm cont} \ \ \ } 
\over {\d A_{x,\m}^{a \ {\rm cont}} } }
= - g_0 a_s^3(D_\l F_{\l\m})^{\rm cont}   \ ,
}}
and we take for the lattice drift force
\eqn\eeaaa{\eqalign{
f_{x,\m, a} = - E_{x,\m, a} S(U) \ .
}}

	The lattice color-electric field operator defined by the
remarkably simple expression \eeaa\ represents the color-electric flux
carried by the link
$(x,\m)$.   It satisfies the Lie algebra commutation relations
\eqn\eeab{\eqalign{
[E_{x,\m, a}, E_{y,\n, b}] = - \d_{x,y} \d_{\m,\n} \ f^{abc} E_{x,\m, c} \ , 
}}
as well as
\eqn\eezz{\eqalign{
[E_{x,\m, a}, U_{y,\n}] &= \d_{x,y} \d_{\m,\n} \ t_a U_{y,\n} \cr
[E_{x,\m, a}, U_{y,\n}^{-1}] &= - \d_{x,y} \d_{\m,\n} \ U_{y,\n}^{-1}t_a\ .
 }}
We may visualize the color-electric field operator $E_{x,\m, a}$ acting on
$S(U)$ as inserting $t_a$ into the plaquettes that contain the link $(x,\m)$
on the left side of the matrix $U_{x,\m}$.  

	The color-electric field operator that
operates by right multiplication of $t_a$,  
\eqn\eead{\eqalign{
{E'}_{x,\m, a}  \equiv
 (U_{x,\m} t_a)_{\a \b} { {\p \ \ \ \ \ \ \ \ \ \ } 
\over { \p(U_{x,\m})_{\a \b} } } \ 
 }}
satisfies the Lie algebra commutation relations with opposite sign
\eqn\eeae{\eqalign{
[{E'}_{x,\m, a}, {E'}_{y,\n, b}] 
= \d_{x,y} \d_{\m,\n} \ f^{abc} {E'}_{x,\m,c}
\ . }}
Since left and right multiplication commute, we have
\eqn\eeaf{\eqalign{
[E_{x,\m, a}, {E'}_{y,\n, b}] = 0 
\ . }}
With 
\eqn\eeag{\eqalign{
U_{x,\m} t_a = (U_{x,\m} t_a U_{x,\m}^{-1}) \ U_{x,\m}
= O_{ba}(U_{x,\m}) \ t_bU_{x,\m} \ ,
}}
where the real orthogonal matrices 
$O_{ba}(U_{x,\m}) = O_{ab}(U_{x,\m}^{-1})$ form the adjoint representation of
the group, the two are related by
\eqn\eeah{\eqalign{
{E'}_{x,\m, a} = O_{ba}(U_{x,\m})E_{x,\m, b} \ ,
}}
and satisfy
\eqn\eeai{\eqalign{
\sum_a {E'}_{x,\m, a}^2 = \sum_a E_{x,\m, a}^2 \ .
}}

	The generator of infinitesimal gauge transformations $G_{x,a}$ is easily
expressed in terms of these operators.  A generic infinitesimal gauge
transformation 
$U_{x,\m} \to g_x^{-1}U_{x,\m}g_{x+\hat{\m}}$, with $g_x = 1 + \omega_x$, is given by
\eqn\eeajn{\eqalign{
\d U_{x,\m} & = - \omega_x U_{x,\m} + U_{x,\m}\omega_{x+\hat{\m}}   \cr
	& = (D_\m \omega)_x U_{x,m} \ ,
}}
where
\eqn\eeajm{\eqalign{
(D_\m \omega)_x \equiv U_{x,\m} \omega_{x+\hat{\m}} U_{x,\m}^{-1}
- \omega_x
= t_a (D_\m \omega)_{x,a} \ .
}}
This defines the lattice gauge-covariant difference $D_\m$ that corresponds
to the continuum gauge-covariant derivative.  It has a simple geometric
meaning, being the difference between $\omega$ at $x+\hat{\m}$ transported
back to $x$, and
$\omega$ at $x$.  This infinitesimal gauge transformation induces the following 
change in a generic function $\CF(U)$:
\eqn\eeaj{\eqalign{
\d \CF & = \CF(U + \d U ) - \CF(U)    \cr
	      & = \sum_{x,\m} (D_\m \omega)_{x,a} \ E_{x,\m,a}\CF     \cr
   &= \sum_x \omega_{x,a} \ G_{x,a}\CF  \ .
}}
This defines the generator of local gauge transformations $G_{x,a}$.  From
\eqn\eeak{\eqalign{
(D_\m \omega)_{x,a} = O_{ab}(U_{x,\m}) \ \omega_{x+\hat{\m},b} 
- \omega_{x,a} 
}}
we obtain
\eqn\eeam{\eqalign{
G_x^a \equiv \sum_\m D_\m^{\dag} E_\m
= \sum_\m ({E'}_{x-\hat{\m},\m}^a - E_{x,\m}^a) , }} 
which is the total flux of color-electric field leaving the site $x$.  As
in the continuum theory, it is the left-hand side of Gauss's law.  One easily
verifies the commutation relations
\eqn\eeal{\eqalign{
[G_x^a, E_{y,\m}^b] &= \d_{x,y} \ f^{abc}E_{y,\m}^c   \cr 
[G_x^a, {E'}_{y,\m}^b] &= \d_{x-\hat{\m},y} \ f^{abc} {E'}_{y,\m}^c   \cr
[G_x^a, G_y^b] &= \d_{x,y} \ f^{abc}G_y^c \ .
}}
Under local gauge transformation, $E_{x,\m}^b$ transforms like
a site variable at $x$ in the adjoint representation, as required for the
gauge-invariance of the transition matrix $\CT(U_{t+1},U_t)$.  Moreover
${E'}_{x,\m}^b$ transforms like a site variable in the adjoint
representation at $x+\hat{\m}$.

	We shall implement a lattice gauge fixing that corresponds to the continuum
gauge fixing described above.  We introduce a minimizing function,
$\CF(U_t)$, and we impose the gauge condition
\eqn\lgfa{\eqalign{
U_{x,t,5} = \exp(A_{x,t,5}) = \exp \Big ( - \e G_{x,t} \CF(U_t)\Big ) \  .  
}}
With this definition, eq. \fdca\ is satisfied, because both $U_{x,t,5}$ and
the gauge-fixing force $f_{x,t,\m}$ are expressed in terms of
variables that live only on the hyperplane $t$.  For the minimizing
function we take a lattice analog of 
$a^{-1}\int d^4x (A^{\rm cont})^2$, for example,
\eqn\lgf{\eqalign{
\CF(U) = 2 a^{-1}\sum_{x,\m} {\rm Re Tr}(1 - U_{x,\m})   \cr
\CF(U) \sim   g_0^2 a^{-1}\int d^4x (A^{\rm cont})^2  \ .  
}}
where $a > 0$ is a gauge parameter (not to be confused with the lattice
units $a_s$ and $a_t$).   This gives the gauge condition
\eqn\lgg{\eqalign{
A_{x,t,5} & = - \e G_x^b\CF(U) \cr
& = - \e a^{-1}\sum_{\m} {\rm Tr}
   \Big ( \ t^b \ (U_{x,\m} - U_{x,\m}^{-1} - U_{x-\hat{\m},\m} +
U_{x-\hat{\m},\m}^{-1}) \Big ) \ .   }}

\subsec{Five-dimensional critical limit and new algorithm}

	The renormalizability of the 5-dimensional continuum theory demonstrated above
controls the critical limit of the 5-dimensional lattice theory just described.
In particular it fixes the dependence of the parameter
$\e = \e(g_0)$ on $g_0$, and thereby provides a new accelerated
Langevin algorithm, as we shall see.  

	The probability for transition from configuration
$U_t$ into the volume element $\prod_{x,\m} dU_{x,t+1,\m}$, where 
$dU_{x,t+1,\m}$
is Haar measure, is given by
\eqn\naa{\eqalign{
\prod_{x,\m} dU_{x,t+1,\m} \CT(U_{t+1},U_t) \ .
}}
From eq. \fdcfa, and invariance of Haar measure under
translation on the group, it follows that the probability for transition
from the configuration $U_t$ to the configuration
\eqn\nab{\eqalign{
U_{x,t+1,\m} =  U_{x,t,5}^{-1} \exp(\e f_{x,t,\m})
U_{x,t,\m}U_{x+\hat{\m},t,5} W_{x,t,\m}  \ 
}}
depends only on the group element $W_{x,t,\m}$,
\eqn\nac{\eqalign{ 
\prod_{x,\m} dU_{x,t+1,\m} \CT(U_{t+1},U_t) =
\CN \prod_{x,\m} \{ dW_{x,t,\m}  \exp\Big ( - \b {\rm Re \ Tr}
(I - W_{x,t,\m})\Big ) \ \}.
}}
Thus to generate a new configuration it is
sufficient to generate the last probability distribution independently for each
group element
$W_{x,t,\m}$, and also independently of the configuration $U_t$, and then
calculate $U_{x,t+1,\m}$ from eq. \nab.  (If one is not interested in
gauge-fixing, one may set
$U_{x,t,5} = 1$.  However the gauge-fixing defined above smooths out
configurations, which may help to accelerate the inversion of the Dirac
operator.)  One recognizes the familiar Langevin algorithm for the SU(N)
group \ukawa\ and \wilson, with $W_{x,t,\m}$ being white noise on the group. 
The new element is that we shall determine $\e = \e(g_0)$ from the perturbative
renormalization group.

	We noted in our discussion of the continuum theory that the time is
renormalized according to $t = Z_t t_r$.  The parameter $\e = a_t/a_s^2$
rescales in the same way as the time $t$, 
\eqn\nafa{\eqalign{
\e = Z_t \e_r \ ,
}}
where $\e_r$ is independent of the ultraviolet cut-off $\L = a_s^{-1}$.
Consequently it satisfies the renormalization-group equation
\eqn\nafb{\eqalign{
\L { {d\ln \e} \over {d\L} } & = \L { {d\ln Z_t} \over {d\L} } \cr
& \equiv \b_t(g_0)   \cr
& = - b_{t,0} g_0^2 + O(g_0^4) \ ,
}}
where $\b_t(g_0)$ is a $\b$-function for the renormalization of the Monte
Carlo time.  The coefficient
$b_{t,0}$ has been calculated for a theory without quarks, \munoz and \okano,
and is independent of the gauge parameter $a$,
\eqn\nafc{\eqalign{ b_{t,0} 
= - {{13 N}\over {3}}{{1}\over {16\pi^2}}
\ . }}
It is convenient to change dependent variable from $\L$ to $g_0$, using 
\eqn\nafd{\eqalign{
\L { {d\ln \e} \over {d\L} }
  = \b(g_0) { {d\ln \e} \over {d g_0} } \ 
}}
where 
\eqn\nafe{\eqalign{
\b(g_0) \equiv \L { {d g_0} \over {d\L} }
}} 
is the usual $\b$-function,
\eqn\naff{\eqalign{
\b(g_0) = - b_0 g_0^3 - b_1 g_0^5 + O(g_0^7) \ .
}}
It has been verified to one-loop level for the 5-dimensional theory without
quarks that
$\b(g_0)$ is the same as in the Faddeev-Popov theory, \munoz
and \okano.  For the theory without
quarks we have
\eqn\naffg{\eqalign{
b_0  = {{11 N}\over {3}} {{1}\over {16 \pi^2}}  \ ,
}}
which gives
\eqn\naff{\eqalign{
{ {d\ln \e} \over {d g_0} } & = { {\b_t(g_0)} \over{\b(g_0)} }  \cr
& = { {b_{t,0}} \over {b_0} } g_0^{-1} + O(g_0)  \ ,
}}
so
\eqn\nafg{\eqalign{
\e = (b_0g_0^2)^{b_{t_0}/(2b_0)} \exp\Big (O(g_0^2)\Big ) \ C \ ,
}}
where $C$ is a constant of integration, and
\eqn\nafh{\eqalign{
{{b_{t,0}}\over {b_0 }} = - {{13}\over {11 }} \ 
}}
for the theory without quarks. The
higher loop corrections are negligible in the critical limit $g_0 \to 0$.  This
allows us to take for the purposes of numerical simulation
\eqn\nafg{\eqalign{
\e = (b_0g_0^2)^{b_{t_0}/(2b_0)} \ C \ ,
}}
and
\eqn\nafh{\eqalign{
\b = (g_0^2 \e)^{-1} = \Big ((g_0^2)^{1 + b_{t_0}/(2b_0)} \
b_0^{b_{t_0}/(2b_0)} \ C\Big )^{-1} \ . }}  
In a theory without quarks, this gives
\eqn\nafha{\eqalign{
\e & = (b_0g_0^2)^{-13/22} \ C   \cr 
\b & = (g_0^2 \e)^{-1} = \Big ((g_0^2)^{9/22} \
b_0^{-13/22} \ C\Big )^{-1} \ .    
}}  

	We note that $\beta$ diverges in the critical limit $g_0 \to 0$, but more
slowly than $1/g_0^2$.  Moreover the time-step in lattice units diverges, $\e =
a_t/s_s^2 \to \infty$ because $b_0$ and $b_{0,t}$ have opposite sign in a
gauge theory.  This provides a highly accelerated algorithm.  However one may
doubt whether this algorithm converges because high wave-number components, $k
\sim a_s^{-1}$ will overshoot and thus may appear to fluctuate wildly.  However
they are not of physical interest in the critical limit.  On the other hand the
modes that are of physical interest 
$k \sim \L_{QCD}$ will not overshoot.  Indeed the real problem is likely to be
that they evolve too slowly as usual, in which case the divergence of the
step-size
$\e(g_0) \to \infty$ as $g_0 \to 0$ is a great asset.  Obviously the algorithm
must be studied in practice before conclusions can be drawn. In order to
apply the algorithm proposed here to a theory with quarks, the
coefficient $b_{t,0}$ must also be calculated with quarks.  

	If one takes $C \to \infty$, one obtains a different, or
non-existent, critical theory from the continuum theory discussed
above. If one takes $C \to 0$, one obtains the continuous-time
stochastic theory discussed in the next section with a Euclidean
spatial cut-off.

	The 5-dimensional critical limit is
approached in the limit $g_0 \to 0$.  The value of $g_0$ determines the
physical correlation length as measured in Euclidean lattice units $a_s$
in the same way that it does in the Faddeev-Popov theory,
\eqn\nafi{\eqalign{
a_s = \L_{QCD}^{-1} (b_0g_0^2)^{-b_1/(2b_0^2)} 
\exp\Big (-{{1}\over {2b_0g_0^2}}\Big ) \ ,
}}
where $\L_{QCD}$ is the standard QCD mass scale, although this has been
verified to one-loop level in the theory without quarks \munoz\ and \okano. 
This formula and the above expression for
$\e(g_0)$ fix the ``time'', $a_t$, for a single Monte Carlo sweep in terms of
$\L_{QCD}$,
\eqn\nafj{\eqalign{
a_t & = \e a_s^2   \cr
a_t & =  \L_{QCD}^{-2} C  
(b_0g_0^2)^{(b_{t,0}b_0 - 2b_1)/(2b_0^2)} \exp\Big (-{{1}\over {b_0g_0^2}}\Big
). }}
In the asymptotic scaling region, auto-correlation
times of physical observables are finite and independent of $g_0$ and $C$
when measured in units of $\L_{QCD}^{-2}$.  The criteria for choosing the value
of $C$ are similar to those for $g_0$.  The auto-correlation time $\tau$ for a
physical observable should be large compared to one sweep-time,  
$\tau >> a_t$, but small compared the total running time $T$, $\tau << T$,
so that the statistical uncertainty, of order $N^{-1/2}$, is small, where 
$N \equiv T / \tau$.

	If it could be established that Monte Carlo auto-correlation times $\tau$ of
physical observables are related to physical Euclidean correlation lengths $L$,
by a universal relation $\tau = K L^2$, where $K$ is the same for all physical
observables, then physical information such mass ratios, could be extracted
directly from Monte-Carlo auto-correlation times.  In this case the mass
spectrum is given by the eigenvalues of the Fokker-Planck hamiltonian,
as explained in sec. 6.3.

	We briefly mention some attractive features of the algorithm proposed
here, which are those of the Langevin algorithm with a time-step in lattice
units that diverges in the critical limit $g_0 \to \infty$.  There is no
accept/reject criterion.  Every updating
$W_{x,t,\m}$ is accepted, even when the (effective) action is non-local as
occurs with dynamical quarks.  Only the derivative of the Euclidean action $S$
appears in 
\eqn\nad{\eqalign{ 
f_{x,t,\m} = - E_{x,\m}S = - E_{x,\m}S_W 
 + \ g_0^2 {\rm Tr} \Big ((\g_\l D_\l + m)^{-1}E_{x,\m}(\g_\n D_\n)\Big ) \ ,
}}
so one does not have to calculate the determinent of the lattice Dirac
operator, but only its inverse.  Moreover it is sufficient to calculate this
inverse only once per sweep, because a whole sweep of the Euclidean lattice,
described by eqs. \nab\ and \nac, corresponds to a single hyperplane of the
5-dimensional lattice with action \nae.  

	Of course other algorithms may be invented that are based on other
discretizations in which the 5-dimensional critical limit is also achived
in the limit $g_0 \to 0$.

\subsec{Lattice Fokker--Planck and Langevin equations}

One would be tempted to formulate a BRST version of  the lattice gauge
theory descrobed in the previous section, which has discrete stochastic time
$t$ and group-valued link variables $U_{x,t,5}$ asociated to the vertical
(time-like) links.  However Neuberger
\neu\ has shown that lattice BRST gauge-fixing fails if the gauge-fixing
function is continuous on the group manifold.  Moreover Testa \testa\ has
provided a counter-example which shows that if one tries to get around
Neuberger's theorem by choosing a gauge-fixing function that is not continuous
on the group manifold, then inconsistencies in the BRST procedure appear in
the form of BRST-exact quantities $sX$ with non-zero expectation-value $<sX>
\neq 0$.  Therefore in order to comply with Neuberger's theorem we shall make
$t = x_5$ continuous while keeping $x_\m$ discrete, so
$A_{x,5}$ is in the Lie algebra whereas $U_{x,\m}$
is in the Lie group, where $\m = 1,...4$.  Gauge-fixing will be of the form
$A_{x,5} = v_x(U)$, so that a quantity in the algebra is fixed instead
of a quantity in the group.  An alternative approach to BRST gauge-fixing in
lattice gauge theory has recently been provided by Baulieu and Schaden 
\Martin.  

	Before developing the BRST formulation of this 5-dimensional lattice theory,
we first consider the stochastic process described by the lattice Fokker-Planck
and Langevin equations with random variable which is a 4-dimensional Euclidean
lattice configuration $U =\{U_{x,\m}\}$, where $U_{x,\m} \in SU(N)$.  Note
that a Euclidean lattice configuration $U$ is a point on a finite product of
Lie-group manifolds and that a continuous stochastic process on Lie-group
manifolds is the probabilistic analog of quantum mechanics on Lie-group
manifolds.  Both allow a well-defined path-integral formulation.

	The continuous-time stochastic process is described by the Fokker-Planck
equation 
\eqn\lfpe{\eqalign{
{ {\p P} \over {\p t} } = - H_{\rm FP}P \ ,   
}} 
where $P = P(U,t)$ is a probability distribution, and the Fokker-Planck
hamiltonian is given by
\eqn\dfpla{\eqalign{
H_{\rm FP} = -  \sum_{x,\m} E_{x,\m}^a\Big ( E_{x,\m}^a
- g_0^{-2}(D_\m A_5)_x^a  + g_0^{-2}E_{x,\m}^a S \Big )     \cr
H_{\rm FP}^{\dag} = -  \sum_x[ \Big( E_{x,\m}^a - g_0^{-2}(E_{x,\m}^a S) \Big)
E_{x,\m}^a  - g_0^{-2}A_{x,5}^a G_x^a ]   \ ,
}}
where 
\eqn\lgfu{\eqalign{
A_{x,5} = - G_x \CF(U) \ ,  
}} 
and $D_\m$ is the lattice gauge-covariant difference, eq.
\eeajm.  In Appendix A we show that this equation describes the limit of the
discrete-time stochastic process defined previous paragraph in which the
time-step approaches zero, $\e = a_t/a_s^2 \to 0$.

	In the previous
paragraph it was conjectured that the auto-correlation times $\tau$ of physical
observables might be related to Euclidean auto-correlation lengths $L$ by the
universal relation $\tau = K L^2$, where $K$ is the same for all
observables.  If this is true, then the mass spectrum is given by the
eigenvalue equation
\eqn\dfplba{\eqalign{
H_{\rm FP} \P & = m^2 \P     \cr
H_{\rm FP}^{\dag} \F & = m^2 \F \ .
}}	
But for the gauge-fixing ``force'', $(D_\m A_5)_x$ in $H_{\rm FP}$, the
equilibrium solution, satisfying $H_{\rm FP}P_{\rm eq} = 0$
would be given by $P_{\rm eq} = \exp(-g_0^{-2}S)$.  We cannot give the exact
equilibrium solution for non-zero gauge-fixing force because it is not
conservative.  However with lattice regularization gauge-fixing may be
dispensed with.  Then a similarity transformation by $\exp[-(g_0^{-2}/2)S]$ is
sufficient to make the the Fokker-Planck hamiltonian hermitian
\eqn\dfplbb{\eqalign{
H'_{\rm FP} = -  \sum_{x,\m} [ \ 
\Big( E_{x,\m}^a - (g_0^{-2}/2)(E_{x,\m}^a S) \Big) \ 
\Big (E_{x,\m}^a  + (g_0^{-2}/2)E_{x,\m}^a S \Big )   \ ]  \ ,
}}
and $H'_{\rm FP} = {H'}_{\rm FP}^{\dag}$.  Moreover the physically relevant
eigenfunctions are gauge-invariant,
$G_x \F' = 0$.  The eigenvalue equation 
\eqn\dfplbc{\eqalign{
H'_{\rm FP} \P' & = m^2 \P',    
}}
is interesting.  It is both
Euclidean- and gauge-invariant.  Moreover the ground-state or vacuum
wave-function is known exactly, $\P'_0 =
\exp[-(g_0^{-2}/2)S]$, so variational calculations of excited-state
eigenvalues are possible.

	The corresponding Langevin equation is obtained from the $\e \to 0$
limit of the stochastic process defined in eqs. \nab\ and \nac.  It is
given by
\eqn\lle{\eqalign{
\dot{U}_{x,\m} {U}_{x,\m}^{-1}
= g_0^{-2}(D_{\m}A_5)_x - g_0^{-2}E_{x,\m} S + \eta_{x,\m}
,   }} 
where $\eta_{x,\m}$ is Gaussian white noise with two-point correlator
\eqn\llea{\eqalign{
<\eta_{x,\m}(s) \eta_{y,\n}(t) > = 2 \d_{x,y} \d_{\m,\n} \d(s-t)
.   }} 

The gauge choice $A_{x,5} = - G_x \CF(U)$ has the global property that 
for the flow determined by the gauge-fixing force alone,
\eqn\lgfb{\eqalign{
\p_5 U_{x,\m}{U}_{x,\m}^{-1} = g_0^{-2} (D_\m A_5)_x ,  
}}
the minimizing function
$\CF$ decreases monotonically
\eqn\lgfc{\eqalign{
\p_5 \CF \leq 0 \ ,
}} 
for we have
\eqn\lgfd{\eqalign{
\p_5 \CF = g_0^{-2} \sum_{x,\m} E_{x,\m} \CF \ D_\m A_5 
 = g_0^{-2} \sum_x G_x \CF \ A_{x,5} = - g_0^{-2} \sum_x G_x \CF \ G_x \CF\ .
}} 
For the minimizing function \lgf\ , this drives all the $U_{x,\m}$ toward
unity.

 \newsec {Five-dimensional topological lattice gauge theory}

\subsec{Lattice BRST operator}

	We determine the lattice BRST operator $s$ which corresponds to the continuum
BRST operator defined previously.  It possesses the symmetries of
the hypercubic lattice.  In order to exhibit these symmetries, we shall, in
this subsection only, use a symmetric 5-dimensional notation, 
whereby $x = \{x_\m\}$ for $\m = 1,...5$ represents lattice sites
(previously denoted $(x,t)$), and we shall also denote link variables by
$U_{xy} \in$ SU(N), with $U_{yx} = (U_{xy})^{-1}$, where $(xy)$ is a pair of
nearest neighbors.

	Consider a generic infinitesimal transformation $U_{xy}$, 
\eqn\lboa{\eqalign{
 U_{xy} + \d U_{xy} 
= (1 + \omega_{xy})U_{xy} 
= U_{xy}(1 + {\omega'}_{xy}) \ , }} 
where $\omega_{xy}$ and 
${\omega'}_{xy} = (U_{xy})^{-1}\omega_{xy}U_{xy}$ are elements of the
Lie algebra, $\omega_{xy} = t_a \omega_{xy}^a$.  We write
$\d = \e s$, 
$\omega_{xy} = \e \Omega_{xy}$ and ${\omega'}_{xy} = \e {\Omega'}_{xy}$,
where $\e$ is an infinitesimal Grassmann variable, and $\Omega_{xy}$
and ${\Omega'}_{xy}$are Lie algebra-valued Grassmann variables, with
\eqn\lboz{\eqalign{{\Omega'}_{xy} = (U_{xy})^{-1}\Omega_{xy}U_{xy} \ . 
}} 
This determines the action of $s$
on $U_{xy}$, 
\eqn\lbob{\eqalign{ sU_{xy} 
= \Omega_{xy}U_{xy} 
= U_{xy}{\Omega'}_{xy} \ . }} 
From the condition $s^2 = 0$, we obtain 
\eqn\lboc{\eqalign{  
s\Omega_{xy} &= \Omega_{xy}^2  \cr
s{\Omega'}_{xy} &= - {\Omega'}_{xy}^2 \ . \cr }} 
We will not write out explicitly further relations for ${\Omega'}_{xy}$ since
they may be obtained for those of $\Omega_{xy}$.  

	As in the continuum theory, we wish to distinguish the infinitesimal gauge
transformations from among all possible transformations of $U_{xy}$. 
  We pose 
$\omega_x = \e c_x$, where $\e$ is an infinitesimal Grassmann variable,
and $c_x$ is a Lie algebra-valued Grassmann variable, so the infinitesimal
gauge transformation defined in the previous section reads
\eqn\lboc{\eqalign{  
\d_g U_{xy} & = \e(-c_xU_{xy} + U_{xy}c_y ) \cr 
         & = \e(-c_x + U_{xy} c_y U_{yx}  )U_{xy} \ , }}
where we have written $U_{yx} \equiv (U_{xy})^{-1}$.  
  We separate this infinitesimal gauge
transformation out of
$\Omega_{x,\m}$, and write
\eqn\lbod{\eqalign{\Omega_{xy} \equiv \P_{xy} -c_x + U_{xy} c_y U_{yx}  
\ , }}
which defines $\P_{xy}$.  The action of $s$ on $U_{xy}$ now reads
\eqn\lbod{\eqalign{  
s U_{xy} = (\P_{xy}-c_x)U_{xy} + U_{xy}c_y \ . }}
We identify the lattice site variable $c_x$ with
the corresponding continuum scalar ghost $c(x)$, and similarly for $\F_x$,
and assign to them the same transformation law as in the continuum case
namely  
\eqn\lbo{\eqalign{  
sc_x &= \F_x - c_x^2  \cr
s\F_x &= - c_x\F_x + \F_xc_x \ . \cr }} 
The action of $s$ on $\P_{xy}$ is obtained from
\eqn\lboe{\eqalign{  
s \P_{xy} = s \Omega_{xy} + sc_x - s(U_{xy} c_y U_{yx}) \ .  }}
This determines the action of $s$ on all the lattice fields and ghosts.

	We write it in the notation in which it will be used, namely with lattice
sites designated by $(x,t)$, where $x$, is a 4-vector $x = \{x_\m\}$ for 
$\m = 1,...4$, and $t = x_5$.  ``Horizontal" links, corresponding to links
in 4-dimensional Euclidean space-time, are designated by
$(x,t,\m)$, and ``vertical" links by $(x,t,5)$.  In this notation $s$ acts
according to 
\eqn\lbof{\eqalign{  
s U_{x,t,\m} & = (\P_{x,t,\m} + (D_\m c)_{x,t}) \ U_{x,t,\m}  \cr
s U_{x,t,5} & = (\P_{x,t,5} + (D_\m c)_{x,t}) \ U_{x,t,5} \cr 
s \P_{x,t,\m} & = - (D_\m \F)_{x,t} 
 - [c_{x,t},\P_{x,t,\m}] + (\P_{x,t,\m})^2  \cr
s \P_{x,t,5} & = - (D_5\F)_{x,t} 
 - [c_{x,t},\P_{x,t,5}] + (\P_{x,t,5})^2  \cr
sc_{x,t} &= \F_{x,t} - (c_{x,t})^2  \cr
s\F_{x,t} &= - [c_{x,t},\F_{x,t}]  \ ,   }}
which satisfies $s^2 = 0$, and where the lattice gauge-covariant difference
operator $D_\m$ is defined in \eeajm.  Under gauge transformation
$\Psi_{x,t,\m}$ and
$\Psi_{x,t,5}$ transform like site variables located at $(x,t)$.  

	We shall shortly take $t = x_5$ to be a continuous variable, while $x$
remains the discrete label of a 4-dimensional lattice site.  In this
case, instead of the link variable $U_{x,t,5}$ in the Lie group, we require a
variable $A_{x,t,5}$ in the Lie algebra.  The BRST operator acts on
$A_{x,t,5}$ and
$\P_{x,t,5}$ as in the continuum theory,
\eqn\lbpa{\eqalign{  
s A_{x,t,5} &= \P_{x,t,5} + D_5 c_{x,t}   \cr
s \P_{x,t,5} &= - D_5 \F_{x,t} - [c_{x,t}, \P_{x,t,5}] \ ,
}}
where $D_5$ is the continuum covariant derivative and $s$ acts  on the
remaining variables as above.

	In terms of the lattice color-electric field operator
$E_{x,\m}$ and generator of local gauge transformations $G_x$ obtained above,
the explicit form of the lattice BRST operator acting on functions
$f = f(U,
\P, c, \F)$ is given by $sf = Qf$, where $Q$ is the differential operator
\eqn\lbrs{\eqalign{
Q \equiv \sum_x \Big( \P_{x,\m}E_{x,\m} + c_x G_x 
&+ \{ -(D_\m \F)_x - [c_x, \P_{x,\m}] + \P_{x,\m}^2 \} 
{ {\p} \over {\p \P_{x,\m} } }   \cr
&+ (\F_x - c_x^2) { {\p} \over {\p c_x } }
- [c_x, \F_x] { {\p} \over {\p \F_x } }
\Big) .   }} 

	The action of $s$ on the anti-ghosts and Lagrange multipliers is defined
simply by
\eqn\lboff{\eqalign{  
s\bar \P _{x,t,\m} &= b' _{x,t,\m } \cr
s\bar \P _{x,t,5} &= b' _{x,t,5 } \cr
s b' _{x,t,\m } &= 0 \cr
s b' _{x,t,5 } &= 0 \cr
s \bar\F_{x,t} &= \bar\eta'_{x,t} \cr
s \bar\eta'_{x,t} &= 0 \ . }}
This is the same as in the continuum theory after the
change of variable
\eqn\ggfcb{\eqalign{  
b' _{x,t,\m} & \equiv b _{x,t,\m} - [c_{x,t},\bar\P_{x,t,\m}] \cr
b' _{x,t,5} & \equiv b _{x,t,5} - [c_{x,t},\bar\P_{x,t,5}] \cr
\bar\eta'_{x,t} & \equiv \bar\eta_{x,t} -[c_{x,t},\bar\F_{x,t}] \ .  }} 
Note that $\F_{x,t}, \bar\F_{x,t}, b_{x,t,\m}$ and $b_{x,t,5}$ are real
variables.

	It is remarkable that the lattice BRST transformation requires only minimal
change from the continuum BRST transformation, even though there is no local
lattice curvature and no lattice Bianchi identity, whereas our starting point
for the continuum transformation was the curvature equation and the Bianchi
identity \gfc.  

\subsec{Five-dimensional topological lattice action}

	The most powerful representation of the continuous stochastic process just
defined is by path integral.  We start with the 5-dimensional topological
lattice action which we write by analogy with the 5-dimensional
continuum action presented above,
\eqn\tla{\eqalign{
I = g_0^{-2} \int dt \sum_x s {\rm Tr} 
\{ \ &  \bar\P_{x,\m}\Big ( (D_5U)_{x,\m}U_{x,\m}^{-1} + E_{x,\m}S(U) 
         + \demi {b'}_{x,\m}\Big )  \cr   
+ & \bar\P_{x,5} ( A_{x,5} + G_x \CF ) \cr
+ & \bar\F_x ( \P_{x,5}  
+ \P_{y,\m} \ E_{y,\m} G_x\CF'  ) \ \} \ ,
}} 
where a sum on $y$ and $\m$ is understood.  Here $\CF(U)$ is the minimizing
function introduced above, and $\CF'(U)$ is the same or another minimizing
function.  

	We expand the above action and obtain
\eqn\tlc{\eqalign{ I = I_1 + I_{A_5} + I_{\P_5} + I_{\P_\m} + I_c + I_\F 
 \ , }}
\eqn\tld{\eqalign{
I_1 & = g_0^{-2} \int dt \sum_x {\rm Tr}\{ \  
  \bar {b'}_{x,\m} \ \Big ( (D_5U)_{x,\m}U_{x,\m}^{-1} + E_{x,\m}S(U) 
         + \demi {b'}_{x,\m}\Big ) \ \} \cr 
I_{A_5} & = g_0^{-2} \int dt \sum_x {\rm Tr} \Big ( \ \bar{b'}_{x,5} \
 ( A_{x,5} + G_x \CF ) \Big ) \cr
I_{\P_5} & =  g_0^{-2} \int dt \sum_x {\rm Tr}\Big (  \bar{\eta'}_x \ 
( \P_{x,5}  + \P_{y,\m} \ E_{y,\m} G_x\CF'  ) \Big )
}}
\eqn\tld{\eqalign{
I_{\P_\m} = - g_0^{-2} \int dt \sum_x {\rm Tr} \Big(
 \bar\P_{x,\m} \ \{ \  & D_5\Big ( \P_{x,\m} + (D_\m c)_x\Big ) 
- D_\m \Big (\P_{x,5} + (D_5 c)_x \Big ) \cr 
& + \Big ( \ \P_{x,\m} + (D_\m c)_x \ , \ 
\p_5U_{x,\m} U_{x,\m}^{-1} - (D_\m A_5)_x \Big )  \cr
& + \Big ( \ \P_{y,\n} + (D_\n c)_y \ \Big ) E_{y,\n} E_{x,\m} S(U) \ \}  \ \Big)
}} 
\eqn\tle{\eqalign{
I_c & = - g_0^{-2} \int dt \sum_x {\rm Tr} \Big( \bar\P_{x,5} \ 
 \{ \ (D_5 c)_x + \P_{x,5} 
+ \Big ( ( D_\m c)_y + \P_{y,\m} \Big ) E_{y,\m} G_x\CF \ \} \ \Big)
}}
\eqn\tlf{\eqalign{
I_\F  =  g_0^{-2} \int dt \sum_x {\rm Tr} \Big( \ \bar\F_x \ 
\{ & - (D_5 \F)_x - [c_x, \P_{x,5}] \cr
& + \Big ( - (D_\m \F)_y - [c_y, \P_{y,\m}] + \P_{y,\m}^2 \Big ) \ 
E_{y,\m} G_x \CF' \ \} \ \Big) \ .
}}
The terms $I_{A_5}$ and $I_{\P_5}$ serve to gauge fix $A_5$ and $\P_5$. 
Indeed upon integration with respect to $\bar{b'}_5$ and $\bar {\eta'}$ they
impose the constraints,
\eqn\tlg{\eqalign{
A_{x,5}^a & = - G_x^a \CF  \cr
\P_{x,5}^a & =  - \P_{y,\m}^b \ E_{y,\m}^b G_x^a\CF' \ .
}}

\subsec{Proof of gauge invariance and absence of Gribov copies.}  

We must show
that the expectation-values of physical observables, namely the Wilson
loops, are independent of the gauge-fixing functions $\CF$ and $\CF'$.  For
this purpose we integrate out $\bar{\eta'}$ and $\bar{\P}_5$, which results
in $\P_5$ being assigned its gauge-fixed value.  We next integrate out the bose
ghosts
$\bar{\F}$ and
$\F$, which appear only in
\eqn\tlga{\eqalign{
Z_{\F} \equiv \int d\F d\bar{\F} \exp(- I_\F) \ .
}}
 The integral on $\bar{\F}$ yields
\eqn\tlh{\eqalign{
Z_{\F} = 
\int \prod_{t,x} d\F_{t,x} \prod_{t,x} \d\Big ( - \p_5 \F_x + (L\F)_x +
f_x\Big ) \ , }}
where $f_x$ is independent of $\F$, and the finite matrix $L$ is defined by
\eqn\tlh{\eqalign{
(L\F)_x^a \equiv - [A_{x,5}, \F_x]^a - \sum_y 
 G_y^b G_x^a \CF' \  \F_y^b \ ,
}}
and we have used $\sum_\m (D_\m^{\dag}E)_{y,\m} = G_y$.
To evaluate $Z_{\F}$, we observe that the stochastic process
starts at some initial time $t = t_0$, which may be taken to be 
$t_0 = - \infty$.   At the initial time, $\F_x(t_0)$, and the other fields
$U, \P, c$, are assigned some definite initial value that is {\it not}
integrated over, for example  $\F_x(t_0) = 0$.  With this initial value,
the differential equation
\eqn\tli{\eqalign{
\p_5 \phi_x - (L\phi)_x = f_x
}}
is equivalent to the integral equation
\eqn\tlj{\eqalign{
\phi_x^b(t) = \int_{t_0}^t du \Big (f_x^b(u) + L_{x,y}^{b,c}(u)\phi_x^c(u)\Big
).   }}
This equation possesses a {\it unique} solution $\phi_x(t)$.  Thus unlike
gauge-fixing on the group, or in a 4-dimensional covariant gauge, there are no
Gribov copies.  

	We shift
$\F$ by $\phi$ which cancels $f_x$ in the path integral for $Z_\F$.  We
also multiply as usual by the formally infinite normalization constant 
$\det \p_5$.  This gives
\eqn\tlk{\eqalign{
Z_\F = 
\int \prod_{t,x} d\F_x(t) \prod_{t,x} \d\Big ( \F_x(t) - (\CL\F)_x(t) \Big ) \ ,
}}
where $\CL =\p_5^{-1}L$ is the integral operator 
\eqn\tll{\eqalign{
(\CL\F)_x^a(t) = \int_{t_0}^\infty du \ \CL_{x,y}^{ab}(t,u) \F_y^b(u),
}}
with kernel
\eqn\tlm{\eqalign{
\CL_{x,y}^{a,b}(t,u) = \theta(t-u) L_{x,y}^{a,b}(u) \ .
}}
This gives
\eqn\tln{\eqalign{
Z_\F = {\det}^{-1}(1 - \CL) 
= \exp \Big(  - {\rm Tr \ ln}(1 - \CL)\Big ) = \exp\Big ( {\rm Tr}\CL + ...\Big
)
\ , }}
Because $\CL_{x,y}(t,u)$ is retarded, $\CL(t,u) = 0$ for $u > t$, only the
tadpole term
\eqn\tlo{\eqalign{
{\rm Tr}\CL & = \int_{t_0}^\infty dt \ \sum_{x,a}\CL_{x,x}^{a,a}(t,t)    \cr
& = \theta(0) \int_{t_0}^\infty dt \sum_{x,a}L_{x,x}^{aa}(t) \ 
}}
survives in the last expansion.  It is natural to assign the
ambiguous expression $\theta(0)$ its mean value $1/2$, which is
consistent with other determinations (see Appendix B), and we obtain 
\eqn\tlp{\eqalign{
Z_\F 
= \exp\Big ( \demi \int_{t_0}^\infty dt \sum_{x,a} L_{x,x}^{a,a}(t)\Big ) \ , 
}}
and so, by eq. \tlh, 
\eqn\tlp{\eqalign{
Z_\F
= \exp\Big ( - \demi \int_{t_0}^\infty dt \sum_{x,a} G_x^aG_x^a \CF'\Big ) \ . 
}}
We emphasize that we have made an exact evaluation of the ghost
determinant, and obtained a purely local contribution to the effective action.

	In the same way we integrate out the Fermi ghosts $\P_\m$ and $\bar{\P}_\m$,
after setting $\P_5$ to its gauge-fixed value.  They only appear in the
factor
\eqn\tlq{\eqalign{
Z_{\P_\m} \equiv \int d\P_\m d\bar{\P}_\m \exp(-I_{\P_\m}) \ .
}}
As before only the tadpole term survives and gives
\eqn\tlqa{\eqalign{
Z_{\P_\m} = 
\exp\Big (\demi \int_{t_0}^\infty dt \sum_{x, a} 
	(\sum_\m E_{x,\m}^aE_{x,\m}^aS + G_x^aG_x^a \CF')\Big ) \ .
}} 
The dependence on $\CF'$ cancels in the product
\eqn\tlr{\eqalign{
Z_{\P_\m} Z_\F = \exp( - I_2) \ ,
}}
where
\eqn\tls{\eqalign{
I_2 \equiv - \demi \int_{t_0}^\infty dt \sum_{x,\m, a} E_{x,\m}^aE_{x,\m}^aS
\ . }}
Thus physical observables are independent of the gauge-fixing
function $\CF'$, as asserted.

	We next show that they are also independent of $\CF$.  The integration on
${b_\m}'$ is effected, which causes
$I_1$ to be changed to ${I_1}'$, given below.  The variable of integration
$c$ may be translated, just as $\F$ was translated above, to cancel the
inhomogeneous term in $I_c$, which changes $I_c$ to ${I_c}'$ given below. 
Putting these factors together we obtain
\eqn\tlt{\eqalign{
Z = \int \prod_{t,x}\Big ( dU_{x,\m}(t) dA_{x,5}(t)
dc_x(t) d\bar{\P}_{x,5}(t) db_{x,5}(t)\Big ) 
\ \exp(-I') \ ,
}}
where 
\eqn\tlu{\eqalign{
I' & = {I_1}' + I_2 + I_{A_5} + {I_c}'   \cr
{I_1}' & =  - \demi  g_0^{-2} \int dt \sum_x {\rm Tr} 
   \ \Big ( (D_5U)_{x,\m}U_{x,\m}^{-1} + E_{x,\m}S(U)\Big )^2  \cr 
{I_c}' & = - g_0^{-2} \int dt \sum_x {\rm Tr} \{ \ \bar\P_{x,5} \ 
\Big ( (D_5 c)_x + ( D_\m c)_y  E_{y,\m} G_x\CF\Big ) \ \} ,
}}
and $I_2$ and $I_{A_5}$ are defined above.  

	We next use standard BRST arguments to show that the
expectation-value of any gauge-invariant observable is independent of the
gauge-fixing function
$\CF$.  Observe first that ${I_1}'$ and $I_2$ are gauge invariant.  They are
therefore invariant under a ``little'' BRST operator $s'$ that acts according
to
\eqn\tlv{\eqalign{ 
s'U_{x,\m} & = (D_\m c)_x U_{x, \m}  \ \ \ \ \
s'A_{x,5} = (D_5 c)_x      \cr
s'c_x & = - c_x^2   \cr
s' \bar{\P}_{x,5} & = b_{x,5} \ \ \ \ \ \ \ \ \ \ \ \ \ \ \  s'b_{x,5} = 0  \ ,
}}
which satisfies $s'^2 = 0$.
Indeed, $s'$ is the usual BRST operator with $\bar{c}_x \to \bar{\P}_{x,5}$
and $b_x \to b_{x,5}$, and it acts on $U_{x,\m}$ and $A_{x,5}$
like an infinitesimal gauge transformation.  Because
${I_1}'$ and $I_2$ are gauge-invariant they are also invariant under $s'$,
\eqn\tlw{\eqalign{
s'{I_1}' = 0;    \ \ \ \ \ \ \ \ \ \  s'I_2 = 0 \ .
}}
Moreover we may express the total action $I'$ as
\eqn\tlx{\eqalign{
I' & = {I_1}' + I_2 + s'J   \ ,
}}
where
\eqn\tlxa{\eqalign{
J  \equiv g_0^{-2} \int dt \sum_x {\rm Tr}\Big (\bar\P_{x,5} \ 
 ( \ A_{x,5} + G_x\CF \ ) \Big ) .
}}
Thus $I'$ is $s'$-invariant, $s'I' = 0$, and moreover the gauge-fixing
function $\CF$ appears only in the term $s'J$ that is $s'$-exact.  By
standard arguments it follows that the expectation-value of any
gauge-invariant observable is independent of $\CF$, as asserted, which
completes the proof.

	We next integrate out the remaining auxiliary
variables $b_5, A_5, \bar{\P}$ and $c$ to obtain a path-integral in terms of
the actual random variables
$U_{x,\m}(t)$ only.  In eq.
\tlt\ , we integrate out $b_{x,5}(t)$ and $A_{x,5}(t)$, which fixes 
$A_{x,5} = - G_x\CF(U)$.  The integral on the Fermi ghosts $c_x(t)$ and
$\bar{\P}_{x,5}(t)$ namely
\eqn\tly{\eqalign{
Z_c & \equiv \int \prod_{t,x}[dc_x(t) d\bar{\P}_{x,5}(t)] 
\ \exp(-{I_c}') \ ,
}} 
may be effected just like the integral on the bose
ghosts $\F_x(t)$ and $\bar{\F}_x(t)$ that was done explicitly above.  Again
only the tadpole term survives and gives
\eqn\tlz{\eqalign{
Z_c & = \exp(-I_3)    \cr
I_3 & = - \demi \int_{t_0}^{\infty} dt \sum_{x,a} G_x^a G_x^a \CF \ .
}}
This yields the desired partition function of the stochastic process in
terms of the random variables $U_{x,\m}(t)$ only,
\eqn\tlza{\eqalign{
Z  = \int \prod_{t,x}dU_{x,\m}(t) \ \exp(-I'') \ , 
}}
where
\eqn\tlzb{\eqalign{
I'' = {I_1}' + I_2 + I_3  
}}
is a local action given by
\eqn\tlzc{\eqalign{
I'' = - \demi \int_{t_0}^{\infty} dt \sum_x \{ \ 
g_0^{-2} {\rm Tr} \Big ( (D_5U)_{x,\m}U_{x,\m}^{-1} + E_{x,\m}S(U)\Big )^2  \cr 
+ E_{x,\m}^aE_{x,\m}^aS + G_x^a G_x^a \CF  \ \} \ .
}}
The last two terms are the famous Ito terms that are
discussed in Appendix B, and we see that they are produced automatically by the
5-dimensional topological action when all auxiliary and
ghost fields are integrated out.  Thus the ghost determinant is
well-defined, and moreover its explicit evaluation gives a local Ito
term to the effective action which comes from the tadpole diagrams only. 
The Ito terms, which involve the second derivative of the action $S$
and of the minimizing function $\CF$ with an over-all minus sign, favor minima
of $S$ and $\CF$ over maxima and saddle points.  This is physically
natural and transparent, so it is somewhat surprising that the tadpole
diagrams, which give the Ito terms with lattice regularization, vanish with
dimensional regularization \danzinn.

\newsec{Conclusion and Perspectives  }

We have seen that the description of the 4-dimensional Yang--Mills theory
in the 5-dimensional framework allows one to obtain a field theory
description that does not suffers from the contradictions of the four
dimensional description. From a physical point of view, we have introduced
in a gauge-invariant way the additional time of the Monte Carlo description
of the quantum theory and unified it with the Euclidean space-time coordinates.
It is of course quite remarkable that this approach permits one to get a fully
consistent lattice formulation of the theory with a natural limit toward a
continuum formulation, while the Gribov ambiguity is just absent. Our work
shows that this ambiguity is merely an artifact of a purely 4-dimensional
description. It is only in five dimensions that one has a globally correct
gauge fixing which is purely local. It thus appears that we can consider the
4-dimensional physical space as a slice in a 5-dimensional manifold, in which
one can express the Yang--Mills theory under the form of a particular
topological field theory. A way to this this schematically is to observe  that
the 4-dimensional action can always be rewritten as the integral of a
topological term in five dimensions:
 \eqn\truc{\eqalign{ \int d^4 x {\cal L} (\phi) =\int d^4 x dt   {{\delta 
{\cal L} (\phi) }\over {\delta \phi}}
 {{d  \phi  }\over {dt}}
 }} Thus, the definition of a quantum field theory amounts to the BRST
invariant gauge-fixing of the action, a task  that we have explained in
detail above and which leads to the solutions of several problems of the
purely 4-dimensional formalism.

	We briefly indicate how our method extends to fermions.  To
obtain the 5-dimensionsal fermionic action of a
4-dimensional Dirac spinor $q$, which automatically extends to a 5-dimensional
spinor, it is now quite natural to introduce a topological BRST operator, and
to enforce the relevant Langevin equation by a BRST-exact action.  Call  $\P_q$
and $\bar\P_q$ the
 commuting  topological ghost and anti-ghost of the  anti-commuting  spinor
$q$, and $b_q$ the anti-commuting Lagrange multiplier field. Then one  has
$sq = \P_q-cq$, $s\P_q = -c\P_q + \F q$, and one may take 
$s \bar\P_q = b_q$, $sb_q = 0$ or 
$s \bar\P_q = b_q - c \bar\P_q $, 
$sb_q = - c b_q  + \F \bar\P_q$.  
In the Euclidean formulation the anti-commuting
anti-quark spinor field $q^{\dag}$ is an independent field, and it has
corresponding commuting topological ghost and anti-ghost
$\P_q^{\dag}$ and $\bar\P_q^{\dag}$, and  Lagrange multiplier field
$b_q^{\dag}$, with $sq^{\dag} = \P_q^{\dag} - q^{\dag}c$, 
$s\P_q^{\dag} = \P_q^{\dag}c - q^{\dag}\F$.  For the anti-ghost one has again
$s \bar\P_q^{\dag} = b_q^{\dag}$, $sb_q^{\dag} = 0$ or 
$s \bar\P_q^{\dag} = b_q^{\dag} + \bar\P_q^{\dag}c $, 
$sb_q^{\dag} = - b_q^{\dag}c - \bar\P_q^{\dag}\F$.  
All these fields are functions of
$x$ and $t$.  We take the topological action,
\eqn\fermions{\eqalign{ I_q  =&\int d^4 x dt   
\ s\big(
\bar \P_q ^\dagger \{   D_5q   - K [ ( \g^\m D_\m + m) q       
+ {a } b_q ] \ \} \big) \cr 
=&\int d^4 x dt   
\ \big( b_q ^\dagger \{   D_5q  - K [ ( \g^\m D_\m +m) q       
+{a }  b_q] \ \}\cr  
& \quad  +\bar \P_q ^\dagger [  D_5\P_q   - K (\g^\m D_\m + m) \P_q] 
+\ldots \big) \ ,
 }}
which must be added to the 5-dimensional action that we have introduced in the
pure Yang--Mills case.  Here $(\g^\m D_\m + m)q =0$ is the Dirac  equation of
motion  of the 4-dimensional theory and $K $ is a kernel \damtsok.  One can
formally prove, by mere Gaussian integrations,  that the equilibrium
distribution is independent of the choice of the kernel $K$ \huffel. Since we
are in the context of a renormalizable theory, it is in fact necessary to
introduce a kernel with canonical dimensional equal to $1$, that is  $K= -
\g^\m D_\m + M$. This is remarkable, since this coincides with the necessity
of having a well-defined Langevin process for spinors, as discussed for
instance in
\huffel. Quite interestingly, power counting would allows a term of the type
$\Gamma^{\m\n} F_{\m\n}$ that is not proportional to the kernel: such term
would contradict the  stochastic interpretation of the 5-dimensional theory
and could be the signal of an anomaly. For vector theories, such term should
not be generated perturbatively.  

Beyond solving the ambiguities which occur in a purely 4-dimensional
approach, our  work   might offer  new perspectives.  We have already
mentioned in sec. 6 the existence of a new algorithm and the conjecture that
the mass spectrum is given by the eigenvalues of the Fokker-Planck
hamiltonian.

 One speculation is the question of
understanding chiral fermions in a lattice formulation.
The topological fermionic action \fermions\ is interesting from this
point of view.  One may consider introducing a $t$-dependent mass term $\m(t)$
in \fermions.  This  might lead to an action
of the type used by Kaplan and Neuberger to solve the chiral fermion problem for
the lattice \kaplan\neubergerr\shamir. In their picture,  one  introduces a
tower of additional fermions, which  can be
combined by Fourier transform into a 5-dimensional fermion.  One  
notices that the equation of motion of  \fermions\ reproduces their
5-dimensional Dirac equation when  $\m(t)$ is a step function. Moreover, if
one gauges the fermion conservation number in the fifth dimension,  which
gives an additional $U(1)$ gauge field $A_{\rm abel}(x,t)$, then $\m(t)$ could
be understood as a special value for the gauge field component
$A_{5{\rm abel}}(x,t)$, similar to a kink that  cannot be set globally
to zero by local gauge-fixing.  Therefore the choice of $\m(t)$ could be
determined by topological considerations.  This issue of seeing whether one
can understand the
 4-dimensional space in which the stochastic process stops as a domain wall
in which everything that remains from the fields of the topological
multiplet of $q$  is a chiral fermion seems to us to be   great interest.

The second perspective  is that of a dual formulation which would  now
involve $2$-form gauge fields.  It is worth indicating the idea here. In the
5-dimensional framework, the   dual of a one-form is  a $2$-form, since
the sum of the degrees of the  curvatures of $A$ and  its dual must equal
$5$.

If one restricts to the abelian part,   one has various 
possibilities for expressing the theory by using arguments as in  \samson. 
 The dual action, up to gauge-fixing and supersymmetric terms, should   
take  the form of a
very simple   5-dimensional  topological term:
\eqn\actionsix{\eqalign{  
 \int _5   B_2\w  d B_2.   
 }} Such a Chern--Simons   type  action  is   interesting, since it can
be gauge-fixed in the framework of the self-duality for 2-form gauge fields
\west. Thus, the question also arises whether  a reliable link can be
established   between the TQFT generated by \actionsix\ and the physical
Yang--Mills theory! 

Let us finally  mention the following observation. Suppose
we start from an action with two abelian two-forms in six dimensions, 
$\int _6  d B_2\w  d\bar B_2$. The field content
that one obtains by dimensional reduction in four dimensions will be made of
$2$ two-forms, $4$ one-forms and $2$ zero-forms. But the  $2$ two-forms are
equivalent by duality to $2$ zero-form in four dimensions. So, we obtain a
field content equivalent to $4$ one-forms and $4$ zero-forms. It is striking
that this is the bosonic  content of the Weinberg-Salam model, with its four
gauge fields and four scalars, which are the components of a complex $SU(2)$
doublet.

\vskip 3em

  \centerline{\bf Acknowledgments}

We are grateful to T.~Banks, M.~Douglas, H.~Neuberger, M.~Schaden, M.~Testa
and J.~Zinn-Justin for discussions. The research of Laurent Baulieu was
partially supported by the NSF under grant PHY-98-02709, partially by RFFI
under grant 96-02-18046 and partially by grant 96-15-96455 for scientific
schools.  Laurent Baulieu wishes to thank Rutgers University, where part of
this work was done, for its very pleasant hospitality. The research of
Daniel Zwanziger was partially supported by the NSF under grant PHY-9900769.

\appendix A{Derivation of lattice Fokker-Planck equation}

	We shall demonstrate that the discrete Markov process defined in sec. 6
is described by the Fokker-Plack equation of sec. 6 in the limit 
$\e \to 0$.  We shall derive the lattice Fokker-Planck hamiltonian
$H_{FP}$ from the transition matrix $\CT(U_{t+1},U_t)$, eq. \fdc\ , in the
same way that the Kogut-Susskind hamiltonian is derived from the transfer
matrix of lattice gauge theory.  We follow here the discussion of Creutz
 \creutz.  We introduce a basis of states labelled by group
elements $|U>$, where $U = \{U_{x,\m}\}$, with operators
\eqn\dfpa{\eqalign{ 
\hat{U}_{x,\m} |U> = U_{x,\m} |U> \ .
}}
We must find an operator $T$ with the property
\eqn\dfpb{\eqalign{ 
<U'|T|U> = \CT(U',U).
}}
To order $\e$ it will have the form $T = 1 - \e H_{FP}$, which will allow
us to identify $H_{FP}$.  

	From eqs. \fdc, \fdcf, \eeaaa\ and \lgg, we have 
\eqn\dfpc{\eqalign{
\CT(U', U) = \CN \exp\Big ( - (\e g_0^2)^{-1} \sum_{x,\m}
{\rm Re \ Tr}(I - {U'}_{x,\m}^{-1}H_{x,\m}U_{x,\m})\Big ) \ ,
}}
Here $H_{x,\m} \in SU(N)$ is given by
\eqn\dfpd{\eqalign{
H_{x,\m} \equiv \exp(-\e v_x) \exp(\e f_{x,\m}) \exp(\e \
U_{x,\m} f_{x+\hat{\m},\m}U_{x,\m}^{-1})
 \ ,}}
and the gauge-fixed value of $A_{x,5}$ appears, namely $A_{x,5} = \e v_x$
where
\eqn\dfpe{\eqalign{
v_x \equiv - G_x \CF \ .
}}
We write
\eqn\dfpf{\eqalign{
H_{x,\m} = \exp(\e k_{x,\m}) \ ,
}}
where $k_{x,\m}$ may be interpreted as the total drift force.  
To order $\e$, it is given by
\eqn\dfpg{\eqalign{
k_{x,\m}\ = f_{x,\m} + (D_\m v)_x \ ,
}}
where $D_{\m}$ is the lattice gauger covariant difference, eq. \eeajm.
It is the sum of the drift force $f_{x,\m} = - E_{x,\m}S$
and a ``gauge-fixing force'' $(D_\m v)$.

	Let $T_0$ be the operator whose matrix elements give the transition
probability with total drift force $k_{x,\m} = 0$, 
\eqn\dfph{\eqalign{
<U'|T_0|U> = \CN \exp\Big ( - (\e g_0^2)^{-1} \sum_{x,\m}
{\rm Re \ Tr}(I - {U'}_{x,\m}^{-1}U_{x,\m})\Big ).
}}
We have
\eqn\dfpi{\eqalign{
<U'|T|U> = <U'|T_0|\exp(\e k)U> , 
}}
where $\exp(\e k)U \equiv \{\exp(\e k_{x,\m})U_{x,\m}\}$.  We insert a
complete set of states, 
\eqn\dfpj{\eqalign{
<U'|T|U> = \int dU'' <U'|T_0|U''><U''|\exp(\e k)U> \ . 
}}
To order $\e$ we have
\eqn\dfpk{\eqalign{
<U''|\exp(\e k)U> & = \d(U'', \exp(\e k)U) = \d(\exp(-\e k)U'', U)  \cr
 & = \Big ( 1 - \e\sum_{x,\m} {E''}_{x,\m}k_{x,\m}(U)\Big )\d(U'', U) ,
}}
where ${E''}_{x,\m}$ is the color-electric field derivative operator defined
in sec. 6, that acts on the variable $U''$.   Let $\hat{E}_{x,\m}$ be the
corresponding quantum mechanical color-electric field operator.  In terms of
operators we have shown that to order $\e$
\eqn\dfpm{\eqalign{
T = T_0 \ \Big (1 - \e \sum_{x,\m}\hat{E}_{x,\m}^ak_{x,\m}^a(\hat{U})\Big ) \ .
}}
The evaluation of $T_0$ may be found in \creutz, with the result
\eqn\dfpl{\eqalign{
T = 1 + \e \sum_{x,m} \{ \hat{E}_{x,\m}^a\Big (g_0^2\hat{E}_{x,\m}^a 
- k_{x,\m}^a(\hat{U})\Big ) \}
\ . }}
Thus with $T = 1 - \e H_{FP}$ and $k_{x,\m} = f_{x,\m} + (D_\m v)_x$,
and 
$f_{x,\m} = - E_{x,\m}S$, we obtain the Fokker-Planck hamiltonian
\eqn\dfpl{\eqalign{
H_{FP} = -  \sum_{x,m} E_{x,\m}^a \Big (g_0^2E_{x,\m}^a
- (D_\m v)_x  + E_{x,\m}S\Big ) \ ,
}}
where $v_x = - G_x\CF$.

\appendix B{Equivalence of topological lattice action and lattice
Fokker-Planck equation}

	We wish to show that the 5-dimensional partition function $Z$, eqs. \tlza\
and \tlzc, that was obtained from the topological lattice action by
integrating out all the auxiliary fields, is a path integral representation
of the solution to the Fokker-Plank equation, 
$\p_5 P = - H_{FP}P$, with Fokker-Planck hamiltonian $H_{FP}$ given in eq.
\dfpl .  

	For this
purpose consider the integral defined for arbitrary $O(U')$ by
\eqn\tafp{\eqalign{
\bar O (U) \equiv &
\int dU' O(U') {\cal K} (U',U)    \cr
 \equiv &\int dU' \ O(U') \ \CN \exp \Big (-(4\e)^{-1}\sum_{x,\m}
(B_{x,\m} - \e k_{x,\m})^2 \Big ) \ ,
}}
where $k_{x,\m} = k_{x,\m}(U)$ is defined in eq. \dfpg, and $B_{x,\m}$ is
defined by
$\exp(t_aB_{x,\m}^a) = {U'}_{x,\m} U_{x,\m}^{-1}$.  It will
be sufficient to show that in the limit $\e \to 0$ this expression
corresponds on the one hand to a discretization of the lattice action \tlzc ,
and on the other hand yields
\eqn\tafpa{\eqalign{
\bar O (U) = (1 - \e H_{\rm FP}^{\dag})O(U) \ .
}}
We change variable of integration
by translation on the group $U' = VU$, with $dU' = dV$.  With 
$V_{x,\m}$ parametrized by $V_{x,\m} = \exp(t_aB_{x,\m}^a)$, we have
\eqn\tafpb{\eqalign{
\bar O (U) = \CN \int dB \r(B) \ O\Big (\exp(B)U\Big ) \ \exp \Big
(-(4\e)^{-1}\sum_{x,\m} (B_{x,\m} - \e k_{x,\m} )^2\Big ) \ ,
}}
where $\r(B)$ is Haar measure.
We again change variable of integration by 
$B_{x,\m} = \e^{\demi}C_{x,\m} + \e k_{x,\m}$, and obtain
\eqn\tafpb{\eqalign{
\bar O (U) = \int dC \ \r(\e^{\demi}C + \e k) \ 
O(\exp(\e^{\demi}C + \e k)U) \ \CN \exp \Big (-\quart\sum_{x,\m}
C_{x,\m}^2\Big ) \ .
}}
We now expand to order $\e$,
\eqn\tafpc{\eqalign{
\r_{x,\m}(\e^{\demi}C + \e k) & = 1 + {\rm const} \times \e C_{x,\m}^2  \cr
O\Big (\exp(\e^{\demi}C + \e k)U\Big ) & = 
 \Big (1 + (\e^{\demi}C_{x,\m}^a + \e k_{x,\m}^a)E_{x,\m}^a
    + {1 \over 2} \e C_{x,\m}^a C_{y,\n}^b E_{x,\m}^a E_{y,\n}^b\Big )O(U),
}}
where a sum over repeated indices is understood.  When this is substituted
into \tafpb, the contribution from the expansion of $\r$ is cancelled by the
normalization constant $\CN$, and one obtains eq. \tafpa\ , with
\eqn\tafpd{\eqalign{
H_{\rm FP}^{\dag} = - \sum_{x,\m}\Big (E_{x,\m} + k_{x,\m}(U)\Big )E_{x,\m} \ ,
}}
as desired.

	We next show that the kernel in \tafp\ corresponds to a discretization of
the path integral with action \tlzc. In the preceding calculation the
drift force $k = k(U)$ depends on $U$ but not on $U'$, whereas the path
integral \tlza\ and \tlzc\ implicitly uses the symmetrized drift force
\eqn\tafpf{\eqalign{
k_s \equiv \demi \Big (k(U) + k(U')\Big ) \ .
}} 
So to compare with the path integral, we express the integral \tafp\ in
terms of
$k_s$ plus a correction which will turn out to be the well-known Ito term.  We
have
\eqn\tafpe{\eqalign{
k(U) = k_s + \demi\Big (k(U) - k(U') \Big) \ ,
}}
so, with $U' = \exp(B)U$, we may write to the order of interest 
\eqn\tafpg{\eqalign{
k(U) = k_s - \demi\sum_{y,\n}B_{y,\n}^aE_{y,\n}^ak \ , 
}}
because the last term gives a contribution of order $\e$ as we shall show. This
gives, to the order of interest,
\eqn\tafph{\eqalign{
\exp\{- (4\e)^{-1}
\sum_{x,\m}\Big (B_{x,\m} - \e k_{x,\m}(U) \Big )^2 \} = 
&   \Big ( 1 - 
\quart\sum_{x,\m,y,\n}B_{x,\m}^bB_{y,\n}^aE_{y,\n}^ak_{x,\m}^b\Big )   \cr 
& \times \exp\Big (- (4\e)^{-1}\sum_{x,\m}(B_{x,\m} - \e k_{s,x,\m})^2 \Big) 
\ . }}
Upon substituting this expression into \tafp, one obtains a Gaussian integral
and Gaussian expectation value
\eqn\tafpi{\eqalign{
<B_{x,\m}^bB_{y,\n}^a> = 2\e \d_{x,y} \d_{\m,\n} \d^{a,b} \ ,
}}
which is indeed of order $\e$, as asserted.  Consequently in the exponent we
may replace $B_{x,\m}^bB_{y,\n}^a$ by its expectation value, and to the order
of interest we have
\eqn\tafpj{\eqalign{
\exp\{- (4\e)^{-1}
\sum_{x,\m}\Big (B_{x,\m} - \e k_{x,\m}(U) \Big )^2 \} \to  
 \exp\{ - \sum_{x,\m}\Big ( & (4\e)^{-1}(B_{x,\m} - \e k_{s,x,\m})^2    \      \cr 
		&  + \demi\e E_{x,\m}^ak_{x,\m}^a\Big ) \} \ , 
}}
the last term being the Ito term.
	Observe that with $U' \to U_{t+1}$ and $U \to U_t$, and 
${U'}_{x,\m}U_{x,\m}^{-1} = \exp(B_{x,\m})$, we have 
$B_{x,\m} \sim \e \dot{U}_{x,\m}U_{x,\m}^{-1}$, so the last
expression may be written formally as
\eqn\tafpk{\eqalign{
  \exp\{ - \e \sum_{x,\m}
\Big ( \quart(\dot{U}_{x,\m}U_{x,\m}^{-1} - \e k_{s,x,\m})^2      
		  +  \demi E_{x,\m}^a k_{x,\m}^a \Big )\} \ . 
}}
We have
\eqn\tafpl{\eqalign{
E_{x,\m}^a k_{x,\m}^a = - E_{x,\m}^a E_{x,\m}^a S - G_x^a G_x^a \CF \ ,
}}
and the last expression corresponds to the action \tlzc, as asserted.

\footatend\vfill\supereject\immediate\closeout\rfile\writestoppt
\baselineskip=14pt\centerline{{\bf References}}\bigskip{\frenchspacing%
\parindent=20pt\escapechar=` \input refs.tmp\vfill\eject}\nonfrenchspacing

\bye